\documentclass[journal,12pt,onecolumn,draftclsnofoot,]{IEEEtran}

\usepackage{amssymb}

\usepackage{amsmath}
\usepackage{stmaryrd}
\usepackage{graphicx,graphics,color,psfrag}
\usepackage{graphicx}
\usepackage{booktabs}
\usepackage{subfigure} 
\usepackage{caption}   
\usepackage{float}
\allowdisplaybreaks
\usepackage{algorithm}
\usepackage{bm}
\usepackage{eufrak}
\usepackage{url}
\usepackage[english]{babel}
\usepackage{algpseudocode}
\usepackage{multirow}
\usepackage{enumerate}
\usepackage{cases}
\usepackage{verbatim}

\usepackage{array}
\usepackage{setspace}
\usepackage{xurl}
\usepackage{placeins}
\usepackage{float}

\begin{document}

\title{Revisiting Trace Norm Minimization for Tensor Tucker Completion: A Direct Multilinear Rank Learning Approach
\thanks{Xueke Tong, Hancheng Zhu and Yik-Chung Wu are with the Department of Electrical and Electronic Engineering, The University of Hong Kong, Hong Kong (e-mail: xktong@eee.hku.hk, hczhu@eee.hku.hk, ycwu@eee.hku.hk).} 
\thanks{Lei Cheng is with the College of Information Science and Electronic Engineering, Zhejiang University, Hangzhou 310027, China, and is also with National Key Laboratory of Underwater Acoustic Technology, Harbin Engineering University, Harbin 150001, China (e-mail: lei\_cheng@zju.edu.cn ).} 
\thanks{Yik-Chung Wu is the corresponding author. This work was supported by the National Natural Science Foundation of China under Grant 62371418.} 
}

\author{Xueke Tong, Hancheng Zhu, Lei Cheng, and Yik-Chung Wu}
\maketitle

\begin{abstract}
To efficiently express tensor data using the Tucker format, a critical task is to minimize the multilinear rank such that the model would not be over-flexible and lead to overfitting. Due to the lack of rank minimization tools in tensor, existing works connect Tucker multilinear rank minimization to trace norm minimization of matrices unfolded from the tensor data. While these formulations try to exploit the common aim of identifying the low-dimensional structure of the tensor and matrix, this paper reveals that existing trace norm-based formulations in Tucker completion are inefficient in multilinear rank minimization. We further propose a new interpretation of Tucker format such that trace norm minimization is applied to the factor matrices of the equivalent representation, rather than some matrices unfolded from tensor data. Based on the newly established problem formulation, a fixed point iteration algorithm is proposed, and its convergence is proved. Numerical results are presented to show that the proposed algorithm exhibits significant improved performance in terms of multilinear rank learning and consequently tensor signal recovery accuracy, compared to existing trace norm based Tucker completion methods.

\end{abstract}






\section{Introduction}

Tensor decompositions are widely used to analyze high-dimensional data,
which have inherent advantages over vector and matrix data representations. They could represent multilinear latent structures in many machine learning applications, such as dimension reduction \cite{QZhaoKernelizationTensor-basedModels,LeOY}, completion of structured data \cite{QZhaoBCPF,LeLei}, face recognition \cite{Faces}, hyperspectral image restoration \cite{R1,R3}, and
snapshot compressive imaging (SCI) reconstruction \cite{R2}. In particular, Canonical Polyadic decomposition (CPD) \cite{RBroPARAFAC} and Tucker decomposition \cite{LRTucker0,LRTucker} are two widely used models \cite{ACichockiNMTF,TamaraGTensorD}.
Among the two models, Tucker
decomposition 
has a higher expressive power 
since it decomposes a tensor into a core tensor with factor matrices multiplied along different modes. CPD can be considered as a special case of Tucker decomposition when the core tensor is in 
a super-diagonal structure \cite{TamaraGTensorD}. 

In the real world, many high-dimensional data have an intrinsic low-rank property, which makes the low-rank Tucker decomposition \cite{ACichockiNMTF,TamaraGTensorD,MixtureOfGaussians} a useful model 
to extract information from 
noisy data. 
On the other hand, in many applications, the observed tensor data may contain a lot of missing values. This gives rise to the Tucker completion problem (i.e., to recover the missing data based on a small amount of observed data and the assumption that the data obey a low-rank Tucker structure)
\cite{LRTC3,NRATC}.


If the dimensions of the Tucker core tensor are known, block coordinate descent (BCD) can be readily used to learn the Tucker model \cite{WTucker}.  Unfortunately, the appropriate dimensions of the Tucker core tensor, also known as the multilinear rank \cite{JBKruskalRank,LDeLathauwerMultilinearSVD}, highly depend on the complexity of the data and are usually not known in practice.
Although incorporating other side information such as smoothness could mitigate tensor model over-fitting \cite{JXue1,JXue2}, when these information is not available, choosing a proper multilinear rank is important.
This leads to a number of existing works trying to learn the multilinear rank together with the Tucker model.  In particular, since minimization of tensor rank is not well studied mathematically compared to matrix rank minimization, most existing works connect the tensor rank minimization problem to some kinds of matrix rank minimization via tensor unfolding. For example, \cite{LRTC3,LRTC} minimizes the trace norm (also called the nuclear norm \cite{MatrixRankMinimization} or Schatten one-norm \cite{GuaranteedMinimumRank}) of the recovered tensor unfolded along different dimensions; \cite{CTNM} proposes to minimize the trace norm of the core tensor unfolded along different dimensions; and \cite{NRATC,SquareDeal} propose to minimize the trace norm of an unfolded and reshaped tensor. 

While the above pioneering works provide some rank lowering effects during Tucker completion, their abilities in estimating the multilinear rank in data is rather limited. This is due to the loss in the process of replacing tensor rank minimization with some forms of unfolded matrix rank minimization. To fill this gap, this paper for the first time reveals an equivalent form of Tucker decomposition, in which it appears in the form of a CPD with each factor matrix having a large number of columns but being low-rank. 

In this noval and new interpretation, multilinear rank is reflected in the rank of the factor matrices, and their minimizations would have a direct effect on the multilinear rank minimization. For the optimization procedure, we employ BCD and accelerated fixed point iteration to derive an iterative algorithm. Despite the problem is non-convex, the proposed aglorithm is shown to be convergence guaranteed with the convergent point being at least a Karush-Kuhn-Tucker (KKT) point. After the model fitting, the multilinear rank of Tucker decomposition 
can be accurately recovered
by performing singular value decomposition (SVD) on the learned factor matrices. 
Simulation results on synthetic and real-world applications (RGB image completion, hyperspectiral image completion, Chemometric data analysis) demonstrate that the proposed low-rank factor matrix Tucker completion (LRFMTC) model provides stable and accurate estimates of multilinear rank and smaller tensor completion error compared to ten state-of-the-art Tucker completion
methods.

The remainder of the paper is organized as follows. 
In Section II, a brief review on Tucker decomposition and the previous related optimization modeling are given. 
In Section III, an equivalent Tucker decomposition format and the corresponding LRFMTC model are presented.  
In Section IV, an optimization algorithm for the proposed model is derived and with its convergence analyzed.
Simulation results and discussions are presented in Section V, and finally conclusions are drawn in Section VI.

\textbf{Notation}: Boldface lowercase and uppercase letters will be used for vectors and matrices, respectively. Tensors are  
written as calligraphic letters. Superscript $T$ denotes transpose. The operator $ \text{Tr} (\boldsymbol A)$ denotes the trace of a matrix $\boldsymbol A$. The symbol $\propto$ represents a linear scalar relationship between two real-valued functions.
The operator $\otimes$ is the Kronecker product, $\odot$ is the Khatri–Rao product, $\circ$ is the outer product, and $\ast$ is the Hadamard product. 
$\boldsymbol I_{M}$ represents the $M \times M$ identity matrix. 
The $i^{th}$ row and the 
$j^{th}$ column of a matrix $\boldsymbol A$ are represented by 
$\boldsymbol A_{i,:}$ and 
$\boldsymbol A_{:,j}$
, respectively. 
The $(i,j,n)^{th}$ element, and the $n^{th}$ matrix atom of a third-order tensor $\mathcal Y$ are represented by $\mathcal Y_{i,j,n}$ and $\mathcal Y_{:,:,n}$, respectively.

\section{Tucker Completion and Rationale of Previous Optimization Modeling}

For a third-order tensor $ \mathcal  X \in \mathbb R ^{I_1 \times I_2 \times I_3}$ with its $(i_1,i_2,i_3)^{th}$ entry denoted by $\mathcal  X_{i_1,i_2,i_3}$, the Tucker decomposition is defined as \cite{TamaraGTensorD}
\begin{align} \label{eq:101}
\mathcal  X = \mathcal G \times_{1} \boldsymbol A^{(1)} 
                                     \times_{2} \boldsymbol A^{(2)}
                                     \times_{3} \boldsymbol A^{(3)},
\end{align}
where $\boldsymbol A^{(1)} \in \mathbb R ^{I_1 \times R_{1}}$, $\boldsymbol A^{(2)} \in \mathbb R ^{I_2 \times R_{2}}$ and $\boldsymbol A^{(3)} \in \mathbb R ^{I_3 \times R_{3}}$ are the mode-1, mode-$2$ and mode-$3$ factor matrices, respectively; $\mathcal G \in \mathbb R ^{R_{1} \times R_{2} \times R_{3}}$ is the core tensor; the symbol $\times_{k}$ denotes the tensor–matrix product along the $k$-mode.
$(R_{1}, R_{2}, R_{3})$ denotes the dimensions of the latent spaces, and the assoicated values leading to the minimum of $\sum_{k=1}^3 R_k$  are called the multilinear rank of tensor $\mathcal  X$. 

Without loss of generality, $\{\boldsymbol A^{(k)} \}_{k=1}^3$ are assumed to be of full column rank. Otherwise, we can always perform SVD $\boldsymbol A^{(k)} = \boldsymbol  \Psi^{(k)} \boldsymbol  \Lambda^{(k)}   {\boldsymbol \Phi^{(k)}}^T$,  with $\boldsymbol \Psi^{(k)}$ now takes the role of factor matrix, and $\boldsymbol \Lambda^{(k)} {\boldsymbol \Phi^{(k)}}^T$ being multiplied with $\mathcal G$ and become part of the core tensor.  This also explains the customary assumption of  
$\boldsymbol A^{(k)}$ having orthogonal columns (i.e., ${\boldsymbol A^{(k)}}^{T} \boldsymbol A^{(k)}=\boldsymbol I_{R_k}$) in many existing works \cite{CTNM,NDSidiropoulosTDSPML}.


Given a partially observed tensor $\mathcal  Y$ and its multilinear rank, we can fit $\mathcal  Y$ to the Tucker model by solving the problem: 
\begin{align} \label{eq:8}
\underset{ \mathcal  G, \{\boldsymbol  A^{(k)}\}}{\min}  & \ ||\mathcal  O \ast (\mathcal  Y - [\![\mathcal  G; \boldsymbol A^{(1)}, \boldsymbol A^{(2)}, \boldsymbol A^{(3)}]\!] ) ||^2_F,  
\end{align}
where $[\![\mathcal  G; \boldsymbol A^{(1)}, \boldsymbol A^{(2)}, \boldsymbol A^{(3)}]\!] = \mathcal G \times_{1} \boldsymbol A^{(1)} \times_{2} \boldsymbol A^{(2)} \times_{3} \boldsymbol A^{(3)}$ is the shorthand notation of the Tucker decomposition. The indices of entries of $\mathcal Y$ being observed are encoded in the tensor $\mathcal O$ where $\mathcal  O_{i_1i_2i_3}=1$ if $\mathcal  Y_{i_1i_2i_3}$ is observed and  $\mathcal  O_{i_1i_2i_3}=0$ otherwise. Equation (\ref{eq:8}) is known as the W-Tucker model \cite{WTucker}, and BCD method can be used to update the parameters 
$\{ \mathcal  G, \boldsymbol A^{(1)}, \boldsymbol A^{(2)},\boldsymbol A^{(3)}\}$
one block at a time.

Like other tensor decompositions, the unfolding operations of Tucker allow a matrix view to the tensor data.
For example, if tensor $\mathcal X$ in (\ref{eq:101})
is unfolded along the first mode, the resulting unfolded matrix is
\begin{align}  \label{eq:102}
\boldsymbol  X_{(1)} &= \boldsymbol A^{(1)} \boldsymbol  G_{(1)} (\boldsymbol A^{(3)} \otimes \boldsymbol A^{(2)})^{T} \in \mathbb R^{I_1 \times I_2 I_3},
\end{align} 
where 
$\boldsymbol G_{(1)}$ 
is the mode-1 unfolded matrix
for the core tensor $\mathcal G$. Similar expressions can be obtained if $\mathcal X$ is unfolded along the second or third mode \cite{TamaraGTensorD,TGKoldaMultilinearOperators}.
Based on the unfolding operations, we could give the definition of $n$-rank of a third-order tensor $\mathcal X$ \cite{TamaraGTensorD}:
\begin{align}  \label{eq:3}
n\text{-rank} ( \mathcal  X) = \big( \text{rank}(\boldsymbol  X_{(1)}), \text{rank}(\boldsymbol  X_{(2)}), \text{rank}(\boldsymbol  X_{(3)})   \big).
\end{align}
It turns out that the $n\text{-rank} ( \mathcal  X)$  and the multilinear rank of $\mathcal  X$ are related, which is given by the following property and proved in Appendix A.

\vbox{}
\noindent
\textbf{PROPERTY 1}. At the minimum value of $\sum_{k=1}^3 R_k$, $\text{rank}(\boldsymbol  X_{(k)}) = R_k$ for all $k$. 



\vbox{}

Inspired by \textbf{Property 1}, the multilinear rank can be estimated by finding the rank of $\boldsymbol  X_{(k)}$, which can be achieved by singular value decomposition (SVD) if $\boldsymbol  X_{(k)}$ is fully observed or low-rank matrix factorization if $\boldsymbol  X_{(k)}$ contains missing values. This is in fact the TREL1 algorithm in \cite{QShi}. After the multilinear ranks are estimated, the original tensor $\mathcal X$ can be recovered by any optimization method, such as W-Tucker \cite{WTucker}.

Finding multilinear ranks and then reconstruct the Tucker tensor using optimization methods involve two steps. However, most existing works put these two steps together. One example is 
the low-$n$-rank tensor completion (LRTC) problem: 
\begin{align} \label{eq:4}
 \underset{\mathcal  X}{\min} &   \sum^{3}_{k=1} \text{rank} (\boldsymbol  X_{(k)})  \ \ \ \  
 \text{s.t.}  \ \mathcal  X_{\boldsymbol \Omega}  =   \mathcal  Y_{\boldsymbol \Omega},               
\end{align}
where $\boldsymbol \Omega$ denotes the indices set of the observed elements in $\mathcal  Y$.
However, as the rank function is nonconvex, the trace norm (also called the nuclear norm)\cite{GuaranteedMnimumRank,FPC},  is commonly used to approximate the rank of matrices, 
with the validity justified theoretically in \cite{GuaranteedMnimumRank,ConvexRelaxation}.
This gives rise to the tensor trace norm minimization problem:
\begin{align} \label{eq:5}
\underset{\mathcal  X}{\min}  & \sum^{3}_{n=1} \alpha_k ||\boldsymbol  X_{(k)}||_*   \ \ \ \ 
 \text{s.t.}  \ \mathcal  X_{\boldsymbol \Omega}  =   \mathcal  Y_{\boldsymbol \Omega}, 
\end{align}
where 
$||\boldsymbol  X_{(k)}||_*$ denotes the trace norm of $\boldsymbol  X_{(k)}$, and $\alpha_k > 0$ controls the relative importance among different trace norms. 

If (\ref{eq:5}) can be solved directly, Tucker completion would be done. Unfortunately, (\ref{eq:5}) is difficult to solve as the unknown variables are coupled in the unfolding along different modes. 
Therefore, existing works solve some forms of surrogate of (\ref{eq:5}) instead \cite{Gandy,Tomioka}. For example,
one can introduce auxiliary variables $\mathcal M_{k} = \mathcal X$, and reformulate (\ref{eq:5}) as
\begin{align} \label{eq:7}
\underset{ \mathcal  X, \mathcal  M_k}{\min}  & \sum^{3}_{k=1} \alpha_k ||\mathcal  M_{k(k)}||_*  \ \ \ \   \text{s.t.}  \  \mathcal  X =   \mathcal  M_k,   k=1,2,3.      
   \ \  \mathcal  X_{\boldsymbol \Omega}  =   \mathcal  Y_{\boldsymbol \Omega}. 
\end{align}
With the constraints $\mathcal  X =   \mathcal  M_k$ moved as penalty terms, 
alternating direction method of multipliers (ADMM) algorithm \cite{ADMM} can be applied to solve (\ref{eq:7}), and this gives the HaLRTC algorithm \cite{LRTC3}.

As $\mathcal  M_k$ in (\ref{eq:7}) is only unfolded along the $k^{th}$ mode, in order to connect (\ref{eq:7}) with matrix completion, one can define $\mathcal  M_{k(k)}$ as matrix $\boldsymbol  M_k$, and (\ref{eq:7}) can be written as
\begin{align} \label{eq:6}
\underset{ \mathcal  X, \boldsymbol  M_k}{\min}  & \sum^{3}_{k=1} \alpha_k ||\boldsymbol  M_k||_* 
       + \frac{\beta_k}{2} || \boldsymbol  X_{(k)} - \boldsymbol  M_k||^2_F
   \ \ \ \ 
 \text{s.t.}  
 \  \mathcal  X_{\boldsymbol \Omega}  =   \mathcal  Y_{\boldsymbol \Omega}, 
\end{align}
where the the constraints $\boldsymbol  X_{(k)} = \boldsymbol  M_k$ is put as a penalty term, with $\beta_k$ being a positive regularization parameter.  
Solving (\ref{eq:6}) in the BCD framework, and each subproblem for $\boldsymbol  M_k$ using matrix completion tool, this results in the SiLRTC algorithm \cite{LRTC3}.



Noticing that the model in (\ref{eq:8}) explicitly considers the Tucker structure but not minimizing the rank, while the models in (\ref{eq:5})-(\ref{eq:6}) consider minimizing the ranks but not exploiting the Tucker structure.  It is not difficult to predict that a better model is to incorporate both ideas, which gives the
core tensor trace norm minimization (CTNM) \cite{CTNM} model:

\begin{align} \label{eq:10}
& \underset{ \mathcal  G, \{\boldsymbol  A^{(k)}\}, \{\boldsymbol V_k\}, \mathcal Z}{\min}   \ \frac{1}{3} \sum^{3}_{k=1}   ||\boldsymbol V_k ||_*  
+ \frac{\lambda}{2} || \mathcal Z - [\![\mathcal  G; \boldsymbol A^{(1)}, \boldsymbol A^{(2)}, \boldsymbol A^{(3)}]\!] ||^2_F
  \nonumber \\
& \ \ \ \ \ \text{s.t.}  \ \boldsymbol G_{(k)} =  \boldsymbol V_k, \ {\boldsymbol A^{(k)}}^{T} \boldsymbol A^{(k)}=\boldsymbol I_{R_k}, \ \mathcal Z_{\boldsymbol \Omega} =  \mathcal  Y_{\boldsymbol \Omega},  
\end{align}
where $\mathcal Z$ and $\{\boldsymbol V_k\}$ are the introduced auxiliary variables, and ${\boldsymbol A^{(k)}}^{T} \boldsymbol A^{(k)}=\boldsymbol I_{R_k}$ denotes the Stiefel manifold \cite{StiefelManifold}.
Again, with the constraints $\boldsymbol G_{(k)} =  \boldsymbol V_k$ put as penalty terms, one can solve (\ref{eq:10}) using ADMM algorithm. 

Note that in models (\ref{eq:7})-(\ref{eq:10}), they involve minimizing the ranks of an unfolded tensor (unfolding of $\mathcal X$ in (\ref{eq:7}) and (\ref{eq:6}), and unfolding of $\mathcal G$ in (\ref{eq:10})),
 which results in the constraints $\mathcal X = \mathcal  M_k$ in (\ref{eq:7}), $\boldsymbol  X_{(k)} = \boldsymbol  M_k$ in (\ref{eq:6}), 
 and %
$\boldsymbol G_{(k)} =  \boldsymbol V_k$ in (\ref{eq:10}).
These constraints have to be handled by penalty terms, which unfortunately means that these constraints are only approximately enforced. Therefore, while (\ref{eq:7})-(\ref{eq:10}) are minimizing the ranks of the auxiliary variables ($\mathcal  M_{k(k)}$ in (\ref{eq:7}), $\boldsymbol  M_k$ in (\ref{eq:6}), and $\boldsymbol V_k$ in (\ref{eq:10})), the low-rankness of the auxilary variables may not be effectively propagated back to the original tensor ($\mathcal X$ in (\ref{eq:7})-(\ref{eq:6}), and $\mathcal G$ in (\ref{eq:10})). This makes the recovered tensor $\mathcal X$ not necessarily low-rank.




A major reason for the appearance of the auxiliary variables in (\ref{eq:7})-(\ref{eq:10}) is that the objective functions involve the same unknown tensor ($\mathcal X$ in (\ref{eq:7})-(\ref{eq:6}), $\mathcal G$ in (\ref{eq:10})) unfolded along different dimensions, which makes direct minimization challenging. In order to avoid the auxiliary variables, 
\cite{SquareDeal} notice that if $\mathcal X$ obeys Tucker structure, 
$\boldsymbol  X_{[j]}:= \text{reshape}(\boldsymbol  X_{(1)}, \prod_{k\leq j} I_{k}, \prod_{k > j}  I_{k}) = ( \boldsymbol A^{(j)} \otimes \boldsymbol A^{(j-1)} \otimes \cdot \cdot \cdot \otimes \boldsymbol A^{(1)} ) \boldsymbol G_{[j]} ( \boldsymbol A^{(3)} \otimes \cdot \cdot \cdot \otimes \boldsymbol A^{(j+1)})$, 
where $j \in \{1,2,3\}$.
By the fact that 
$\text{rank}(\boldsymbol P\otimes \boldsymbol Q)=\text{rank}(\boldsymbol P)\text{rank}(\boldsymbol G)$,  
minimizing the rank of $\boldsymbol  X_{[j]}$ would minimize the $\text{rank}(\boldsymbol A^{(j)})\text{rank}(\boldsymbol A^{(j-1)})\cdot \cdot \cdot \text{rank}(\boldsymbol A^{(1)})$ or $\text{rank}(\boldsymbol A^{(3)}) \cdot \cdot \cdot \text{rank}(\boldsymbol A^{(j+1)})$. Consequently, \cite{SquareDeal} proposes to minimize the trace norm of $\boldsymbol  X_{[j]}$:
\begin{align} \label{eq:505}
\underset{\mathcal  X}{\min}  &   ||\boldsymbol  X_{[j]}||_*  \ \ \ \ 
 \text{s.t.}   
\ \mathcal  X_{\boldsymbol \Omega}  =   \mathcal  Y_{\boldsymbol \Omega}. 
\end{align}  
Since minimization of (\ref{eq:505}) can only minimize the rank of $\{\boldsymbol A^{(k)}\}_{k=1}^j$ or $\{\boldsymbol A^{(k)}\}_{k=j+1}^3$, to maximize its effect, \cite{SquareDeal} suggests that 
$j$ is chosen from $\{1,2,3\}$ to make $\prod_{k\leq j} I_{k}$ as close to $\prod_{k > j} I_{k}$ as possible. A recent work \cite{NRATC} further improved (\ref{eq:505}) by applying a concave function \cite{NonConvex} on the singular values of $\boldsymbol  X_{[j]}$ before adding them together (note that trace norm is simply sum of singular values).  The resultant problem is then solved by the proximal linearized minimization (PLM) method.  While both \cite{SquareDeal} and \cite{NRATC} avoid the use of auxiliary variables, 
due to the definition of $\boldsymbol  X_{[j]}$, they only minimize part of the multilinear rank.


From the above discussions, it is clear that none of the existing works solve (\ref{eq:5}) directly, which leads to the recovered tensor not having the best low-rank structure.
In the following, we reveal a new equivalent form of Tucker model, in which the low-rankness is reflected in independent factor matrices. 
This new model not only avoids the use of auxiliary variables, but also minimizes the multilinear rank directly.
These useful properties facilitate the subsequent rank minimization and missing data recovery.

\section{Low-rank Factor Matrices Based Tucker Trace Norm Minimization}

To unveil the equivalent model, 
we represent the core tensor 
$\mathcal G$ by a high-rank CPD: 
$\mathcal G = [\![\boldsymbol \Xi^{(1)}, \boldsymbol \Xi^{(2)}, \boldsymbol \Xi^{(3)}]\!] = \sum_{l=1}^{L} {\boldsymbol \Xi}^{(1)}_{:,l} \circ {\boldsymbol \Xi}^{(2)}_{:,l} \circ {\boldsymbol \Xi}^{(3)}_{:,l} $
where 
$\boldsymbol \Xi^{(1)} \in \mathbb R ^{R_{1} \times L}$, 
$\boldsymbol \Xi^{(2)} \in \mathbb R ^{R_{2} \times L}$, 
$\boldsymbol \Xi^{(3)} \in \mathbb R ^{R_{3} \times L}$. 
As any arbitrary third-order tensor could be represented by a CPD of finite rank \cite{NDSidiropoulosTDSPML}, choosing $L$ large enough would model any Tucker core $\mathcal G$ (If $L$ is small, the modeling capability would be reduced).
With the 
core tensor represented by a CPD, 
the Tucker model is given by
\begin{align}  \label{eq:103}
\mathcal  X &= [\![\boldsymbol \Xi^{(1)}, \boldsymbol \Xi^{(2)}, \boldsymbol \Xi^{(3)}]\!] \times_{1} \boldsymbol A^{(1)}
                                       \times_{2} \boldsymbol A^{(2)}
                                       \times_{3} \boldsymbol A^{(3)}.                     
\end{align}

%

Noticing that unfolding  $\mathcal G = [\![\boldsymbol \Xi^{(1)}, \boldsymbol \Xi^{(2)}, \boldsymbol \Xi^{(3)}]\!] $ along the first mode gives $\boldsymbol  G_{(1)} = \boldsymbol  \Xi^{(1)} (\boldsymbol \Xi^{(3)} \odot \boldsymbol \Xi^{(2)})^{T}$ \cite{TamaraGTensorD} and making use of (\ref{eq:102}), we have  
\begin{align}   \label{eq:104}
\boldsymbol  X_{(1)} &= \boldsymbol A^{(1)}  \boldsymbol  \Xi^{(1)} (\boldsymbol \Xi^{(3)} \odot \boldsymbol \Xi^{(2)})^{T} 
  (\boldsymbol A^{(3)} \otimes \boldsymbol A^{(2)})^{T} \nonumber \\
   &= \boldsymbol A^{(1)}  \boldsymbol  \Xi^{(1)}  
   (\boldsymbol A^{(3)} \boldsymbol \Xi^{(3)} \odot \boldsymbol A^{(2)} \boldsymbol \Xi^{(2)})^{T},
\end{align}
where the last line is due to a known property of Khatri–Rao product \cite{TamaraGTensorD}.
Folding (\ref{eq:104}) back to a 3D tensor gives 
$\mathcal X=[\![\boldsymbol A^{(1)} \boldsymbol \Xi^{(1)}, \boldsymbol A^{(2)} \boldsymbol \Xi^{(2)}, \boldsymbol A^{(3)} \boldsymbol \Xi^{(3)}]\!]$.  

For notational simplicity, if we define $\boldsymbol B^{(k)}=\boldsymbol A^{(k)} \boldsymbol \Xi^{(k)}$ for $k=1, 2, 3$, (\ref{eq:103}) can be rewritten as  
$\mathcal X=[\![\boldsymbol B^{(1)},\boldsymbol B^{(2)},\boldsymbol B^{(3)}]\!] $. 
As $\boldsymbol B^{(1)}$, $\boldsymbol B^{(2)}$, and $\boldsymbol B^{(3)}$ are constructed from multiplication of two matrices, and $R_1$, $R_2$, and $R_3$ are smaller than $I_1$, $I_2$, $I_3$ and $L$, this result reveals that Tucker decomposition is actually equivalent to a CPD if the factor matrices $\boldsymbol B^{(1)}$, $\boldsymbol B^{(2)}$, $\boldsymbol B^{(3)}$ have a large number of columns while being low-rank.

Before we present the optimization formulation for learning $\boldsymbol B^{(1)}$, $\boldsymbol B^{(2)}$, $\boldsymbol B^{(3)}$ with low-rank structure, let us see how we can recover the Tucker structure if we have $\boldsymbol B^{(1)}$, $\boldsymbol B^{(2)}$, $\boldsymbol B^{(3)}$. 
In particular, using
SVD: $\boldsymbol B^{(k)}=\boldsymbol U^{(k)} \boldsymbol D^{(k)} {\boldsymbol V^{(k)}}^T$ for 
$k=1, 2, 3$,  
in $\mathcal X = [\![\boldsymbol B^{(1)}, \boldsymbol B^{(2)}, \boldsymbol B^{(3)}]\!]$, we obtain 
$\mathcal X=[\![\boldsymbol U^{(1)} \boldsymbol D^{(1)}  {\boldsymbol V^{(1)}}^T, \boldsymbol U^{(2)} \boldsymbol D^{(2)}  {\boldsymbol V^{(2)}}^T, $ $\boldsymbol U^{(3)} \boldsymbol D^{(3)}  {\boldsymbol V^{(3)}}^T]\!]$.  Using a reverse logic of $(\ref{eq:104})$, it can be easily shown that $\mathcal X=[\![\boldsymbol D^{(1)}  {\boldsymbol V^{(1)}}^T, \boldsymbol D^{(2)}  {\boldsymbol V^{(2)}}^T, \boldsymbol D^{(3)}  {\boldsymbol V^{(3)}}^T]\!] \times_1 \boldsymbol U^{(1)} \times_2 \boldsymbol U^{(2)} \times_3 \boldsymbol U^{(3)}$.  This recovers the core tensor as $[\![\boldsymbol D^{(1)}  {\boldsymbol V^{(1)}}^T, \boldsymbol D^{(2)}  {\boldsymbol V^{(2)}}^T, $ $\boldsymbol D^{(3)}  {\boldsymbol V^{(3)}}^T]\!]$, and the number of non-zero elements in $\boldsymbol D^{(1)}$, $\boldsymbol D^{(2)}$ and $\boldsymbol D^{(3)}$ give the estimate of $R_1$, $R_2$ and $R_3$, respectively.
We summarize the above observations in the following property.

\vbox{}
\noindent
\textbf{PROPERTY 2}.  The Tucker decomposition $\mathcal X=\mathcal G\times_1 \boldsymbol A^{(1)} \times_2 \boldsymbol A^{(2)} \times_3 \boldsymbol A^{(3)}$ can be expressed as a CPD $\mathcal X = [\![\boldsymbol B^{(1)},\boldsymbol B^{(2)}, \boldsymbol B^{(3)}]\!]$ with factor matrices $\{ \boldsymbol B^{(k)}\}_{k=1}^3$ having a large number of columns while being low-rank.  Given $\mathcal G$, $\{ \boldsymbol A^{(k)}\}_{k=1}^3$, we can construct $\boldsymbol B^{(k)} = \boldsymbol A^{(k)} \boldsymbol \Xi^{(k)}$, where $\boldsymbol \Xi^{(k)}$ is the CPD factor matrix of the Tucker core $\mathcal G$.  On the other hand, given $\{ \boldsymbol B^{(k)}\}_{k=1}^3$, we can perform SVD on $\{\boldsymbol B^{(k)} = \boldsymbol U^{(k)} \boldsymbol D^{(k)} {\boldsymbol V^{(k)}}^T \}_{k=1}^3$ and recover $\mathcal G=[\![\boldsymbol D^{(1)} {\boldsymbol V^{(1)}}^T , \boldsymbol D^{(2)}$ $ {\boldsymbol V^{(2)}}^T , \boldsymbol D^{(3)} {\boldsymbol V^{(3)}}^T ]\!]$, and $\{ \boldsymbol A^{(k)} = \boldsymbol U^{(k)} \}_{k=1}^3$. 

\vbox{}
Now, consider an observed tensor $\mathcal Y$ with missing data
\begin{align} \label{eq:11}
\mathcal O \ast \mathcal Y= \mathcal O \ast ([\![\boldsymbol B^{(1)},\boldsymbol B^{(2)},\boldsymbol B^{(3)}]\!] + \mathcal W),
\end{align}
where  $\mathcal W \in \mathbb R ^{I_1 \times I_2 \times I_3}$ represents the Gaussian noise,  with each element $\mathcal W_{i_1,i_2,i_3}$ follows independently a zero-mean Gaussian distribution with the same variance.
In order to learn $\boldsymbol B^{(k)}$ with low-rank structure, we minimizes the rank of $\boldsymbol B^{(k)}$ together with the error in fitting the observations. If trace norm is used to approximate the rank function, $\boldsymbol B^{(k)}$ can be learnt by solving
\begin{align}     \label{eq:13}
\underset{ \{\boldsymbol B^{(k)}\} }{\min} & \alpha \sum^{3}_{k=1}|| \boldsymbol B^{(k)}||_{*}     
   + \frac{1}{2} \left \|  
 \Big[ \mathcal Y - \sum^{L}_{l=1} \left( \boldsymbol B^{(1)}_{:,l} \circ \boldsymbol B^{(2)}_{:,l} \circ \boldsymbol B^{(3)}_{:,l} \right)  \Big]  \ast  \mathcal O
                       \right\|^{2}_F,
\end{align}
where $\alpha$ 
is a non-negative regularization parameter,
which balances the relative importance of the low-rankness of the recovered $\{ \boldsymbol B^{(k)} \}^3_{k=1}$ and the fitting error of data $\mathcal Y$. 
In particular, when $\alpha$ is large, the estimated tensor tends to have lower multilinear ranks at the expense of increased fitting error. When $\alpha$ is small, the estimated tensor would have a better fit to the data, but the recovered Tucker format may have a higher rank. In addition, if 
the $L_2$ norm in (14) is replaced by $L_1$ norm, the formulation can handle outliers, but such modification is left as a future work.

Notice that PROPERTY 2 reveals a new interpretation of Tucker format, rather
than a new decomposition model. We need this new interpretation of Tucker
because in the conventional optimization of Tucker model, auxiliary variables
are required to decouple the sharing of the same entries along different unfolding directions, which leads to ineffcient learning of the Tucker multilinear ranks. With the new interpretation of Tucker in PROPERTY 2, the equivalent factor matrices $\{ \boldsymbol B^{(k)} \}^3_{k=1}$ are decoupled naturally and each is related to one of the multilinear ranks. This leads to the auxiliary variable-free formulation in (14).  
Although decomposing the Tucker core with CPD exist \cite{R6,R7}, this paper is the first work to establish the relationship between such model and Tucker multilinear rank estimation.
Furthermore, (14) minimizes the rank of factor matrices $\{ \boldsymbol B^{(k)} \}^3_{k=1}$ in the equivalent representation, rather than the rank of factor matrices $\{ \boldsymbol A^{(k)} \}^3_{k=1}$ in the original representation of Tucker in (11) since
according to PROPERTY 2, the ranks of factors $\{ \boldsymbol B^{(k)} \}^3_{k=1}$ are the columns numbers of $\{ \boldsymbol A^{(k)} \}^3_{k=1}$, which are the multilinear ranks of the Tucker representation (please refer to eq. (1)). Therefore, by minimizing the ranks of $\{ \boldsymbol B^{(k)} \}^3_{k=1}$, we could achieve
the multilinear rank minimization of the Tucker decomposition.

While (\ref{eq:13}) also employs trace norm for matrix rank minimization, it
is largely different from (\ref{eq:7})-(\ref{eq:505}).  In particular, there is no unfolding operation and auxiliary variables in (\ref{eq:13}).  Furthermore, the low-rank property is enforced in each of the $\boldsymbol B^{(k)}$. This makes the recovered 
%
$\mathcal X = [\![\boldsymbol B^{(1)},\boldsymbol B^{(2)},\boldsymbol B^{(3)}]\!]$ also low-rank,
and the rank of $\{ \boldsymbol B^{(k)} \}_{k=1}^3$ directly gives the multilinear rank of the Tucker decomposition.
In fact, one can replace the trace norm in (\ref{eq:13}) with other low-rank regularizers such as spectral norm \cite{SpectralNorm},
truncated nuclear norm \cite{TruncatedNN}, weighted nuclear norm \cite{WeightedNN},
nuclear-$l_1$-norm \cite{nuclearL1norm},  
log-sum form \cite{KBR},
Schatten-p norm \cite{schattenPnorm} 
and weighted t-Schatten-p quasi-norm \cite{schattenPQnorm}. We choose the trace norm due to its simplicity and easy comparison with existing works. 
The significance of (\ref{eq:13}) is that the newly discovered representation of Tucker format gives us concrete suggestion that the low-rankness should be imposed on the factor matrices $\{ \boldsymbol B^{(k)} \}^3_{k=1}$, rather than some form of unfolded matrices.
Since we are minimizing the ranks of the factor matrices, we term the proposed method as the Low-Rank Factor Matrix for Tucker Completion (LRFMTC).

While minimizing factor matrix ranks in a CPD has also been proposed in \cite{YLFactorMatrix}, the motivation and argument to derive such formulation are completely different. In particular, \cite{YLFactorMatrix} does not connect such formulation to the minimization of the multilinear rank in Tucker. It simply relaxes the mode-$n$ rank minimization in a CPD completion problem to factor matrix rank minimization.
Moreover, with the theoretical equivalence guaranteed when the columns number of the CPD representation of the Tucker core ($L$) is large enough, we provide a guideline on how to choose $L$ in practical implementation, and this guideline does not exist in \cite{YLFactorMatrix}. If we set the parameter $L$ according to \cite{YLFactorMatrix}, the Tucker core will not be flexible enough to represent arbitrary tensor data obeying the Tucker format.
Furthermore, the algorithm proposed in \cite{YLFactorMatrix} for solving such factor matrix ranks minimization still makes use of auxiliary variables, which unfortunately suffers from ineffective low-rankness propagation from the auxiliary variables to the factor matrices.

\section{Auxiliary Variables-free Optimization Algorithm}  

In this section, we use block coordinate descent (BCD) method to solve problem (\ref{eq:13}), which optimizes one $\boldsymbol B^{(k)}$ at a time while holding $\{\boldsymbol B^{(h)}\}_{h \neq k}$ at the last optimized value.  
While employing BCD seems to be a straightforward choice, this is possible due to the new auxiliary variable-free problem formulation (\ref{eq:13}). In contrast, most existing methods cannot simply employ alternating minimization due to the involvement of auxiliary variables.
From problem (\ref{eq:13}), the 
subproblem with respect to $\boldsymbol B^{(k)}$ is given by 
\begin{align}  \label{eq:14}
\boldsymbol B^{(k)} \gets \arg & \underset{\boldsymbol B^{(k)}}{ \min}   \Bigg\{ \alpha ||\boldsymbol B^{(k)}||_{*} 
                      + \frac{1}{2}  \left \| 
                     \bigg[  \boldsymbol Y_{(k)}  -  
                       \boldsymbol B^{(k)} \Big( \underset{h \ne k}{\bigodot}  \boldsymbol B^{(h)} \Big)^{T} \bigg] \ast  \boldsymbol O_{(k)}
                        \right\|^{2}_F \Bigg\},   
\end{align}  
where $ \boldsymbol Y_{(k)}$ and $ \boldsymbol O_{(k)}$ are the unfolded matrices of tensor $\mathcal Y$ and $\mathcal O$ respectively along the $k^{th}$ mode. In (\ref{eq:14}), all the $\{\boldsymbol B^{(h)}\}_{h \neq k}$ are fixed and
$ \underset{h \ne k}{\bigodot}  \boldsymbol B^{(h)}  =  \boldsymbol B^{(3)} \odot \cdot \cdot \cdot \odot  \boldsymbol B^{(k+1)} \odot  \boldsymbol B^{(k-1)} \odot  \cdot \cdot \cdot  \odot    \boldsymbol B^{(1)} $.


Subproblem (\ref{eq:14}) is a matrix rank minimization problem, and there are many available methods for solving it.  For example, one can reformulate (\ref{eq:14}) 
as a semidefinite programming (SDP) \cite{SDP} or a second-order cone programming (SCOP) \cite{SCOP} problem, 
and then solve it by convex optimization tools. However, solving SDP or SCOP has a very high computational complexity, 
and could not 
handle matrix size more than 
100 $\times$ 100. 
For large-scale matrices, 
fixed point iteration \cite{FPC} is a
lower-complexity solution, and we will illustrate it in the following.


In particular, following the framework of fixed point iteration \cite{FPC}, it is derived  
in Appendix B that the algorithm for solving (\ref{eq:14}) is composed of iterative execution of (over iteration index $t$): 
\begin{align}      \label{eq:201} 
&\boldsymbol Z^t_{k} = {\boldsymbol B^{(k)}}^t  
- \tau_k \bigg[ \bigg( {\boldsymbol B^{(k)}}^t \Big( \underset{h \ne k}{\bigodot}  \boldsymbol B^{(h)} \Big)^{T}
 -  \boldsymbol Y_{(k)} \bigg) \ast \boldsymbol O_{(k)} \bigg]  \Big( \underset{h \ne k}{\bigodot}  \boldsymbol B^{(h)} \Big)      \nonumber \\
&{\boldsymbol B^{(k)}}^{t+1} = D_{\tau_k \alpha} (\boldsymbol Z^t_k),
\end{align}
where $D_{\tau_k \alpha}(\boldsymbol X)$ is the singular value shrinkage operator 
\cite{FPC}, 
$\tau_k \in \Big(0, 2/ $ $ \lambda_{max}  \big( {\oast}_{h \ne k}   ({\boldsymbol B^{(h)}}^T \boldsymbol B^{(h)} ) \big) \Big)$ with $ {\oast}_{h \ne k}  ( {\boldsymbol B^{(h)}}^T \boldsymbol B^{(h)} ) = ( {\boldsymbol B^{(3)}}^T \boldsymbol B^{(3)} ) \ast \cdot \cdot \cdot \ast  ( {\boldsymbol B^{(k+1)}}^T $ $\cdot \boldsymbol B^{(k+1)} ) \ast  ( {\boldsymbol B^{(k-1)}}^T \boldsymbol B^{(k-1)} ) \ast \cdot \cdot \cdot  \ast    ( {\boldsymbol B^{(1)}}^T \boldsymbol B^{(1)} )$,
and $\lambda_{max}(\cdot)$ denotes the maximum eigenvalue of a square matrix.

However, 
since this fixed point iteration is a special case of the iterative shrinkage thresholding framework (Section 3 in~\cite{ABeck}), its convergence rate is therein ${\mathbf{O}}\left( {\frac{1}{t}} \right)$.
%
To accelerate this fixed point iteration algorithm, an extrapolation term can be added and the details are given as \begin{align}      \label{eq:20111} 
&\boldsymbol Z^t_{k} = {\boldsymbol M^{(k)}}^t  
- \tau_k \bigg[ \bigg( {\boldsymbol M^{(k)}}^t \Big( \underset{h \ne k}{\bigodot}  \boldsymbol B^{(h)} \Big)^{T}
 -  \boldsymbol Y_{(k)} \bigg) \ast \boldsymbol O_{(k)} \bigg]  \Big( \underset{h \ne k}{\bigodot}  \boldsymbol B^{(h)} \Big),      \nonumber \\
&{\boldsymbol B^{(k)}}^{t+1} = D_{\tau_k \alpha} (\boldsymbol Z^t_k), \nonumber \\
& u_{k}^{t+1} = \frac{1+\sqrt{1+4(u_k^t)^2}}{2},\nonumber \\
& {\boldsymbol M^{(k)}}^{t+1} = {\boldsymbol B^{(k)}}^{t+1} + \frac{u_{k}^{t}-1}{u_{k}^{t+1}} ({\boldsymbol B^{(k)}}^{t+1} - {\boldsymbol B^{(k)}}^{t}),
\end{align}
with the initial value $u_{k}^0 = 1$ and ${\boldsymbol M^{(k)}}^0 = {\boldsymbol B^{(k)}}^0$.

The key insight of this fast fixed point iteration is that directly setting ${\boldsymbol M^{(k)}}^{t+1} = {\boldsymbol B^{(k)}}^{t+1}$ is too conservative. Therefore, an extrapolation term $({\boldsymbol B^{(k)}}^{t+1} - {\boldsymbol B^{(k)}}^{t})$ is added to make the iterative point ${\boldsymbol M^{(k)}}^{t+1}$ move further (if we drop the extrapolation term, the fast fixed point iteration reduces to the basic fixed point iteration (\ref{eq:201})).  However, the weighting of this extrapolation term should not be too large;
otherwise it is easy to miss the optimal solution. Hence, a monotonically increasing sequence $\{ u_{k}^{t} \}_{t \in \mathbb N}$, which can be viewed as a damping system \cite{ADiff}, is introduced to adjust this ratio. At the beginning, $u_{k}^t$ is small such that over-damping is used to push the solution point ${\boldsymbol M^{(k)}}^{t+1}$ forward. As $t$ increases, $u_{k}^t$ becomes larger, which corresponds to under-damping for pulling the solution ${\boldsymbol M^{(k)}}^{t+1}$ towards the optimal point. According
to \cite{ABeck}  the convergence speed of this accelerated iteration is $\mathbf O (\frac{1}{t^2})$.




With the inner iteration for updating a particular block $\boldsymbol B^{(k)}$ using (\ref{eq:20111}), and
the outer iterations of BCD for alternatively updating different blocks $k=1,2,3$, the procedure of the algorithm is summarized in Algorithm 1. The initialization, stopping criterion, and complexity analysis of Algorithm 1 are given in Appendix C.



\begin{table}[]
\centering
\scalebox{1}{
\begin{tabular}{l}
\hline
\textbf{Algorithm 1}: Tucker completion using LRFMTC 
\\     \hline
\hspace*{0.2cm} \textbf{Input}: a third-order noisy tensor $\mathcal Y$, observation index $\mathcal O$, $\alpha$  \\
\hspace*{0.2cm} \textbf{Initialization}:  ${\boldsymbol B^{(1)}}$, ${\boldsymbol B^{(2)}}$, ${\boldsymbol B^{(3)}}$\\
\hspace*{0.2cm} \textbf{repeat}:     \\
\hspace*{0.4cm} for $k = 1,2,3$        \\
\hspace*{0.6cm} \textbf{repeat}:                                    \\
\hspace*{0.8cm}  Update $\boldsymbol B^{(k)}$ using (\ref{eq:20111}) with$\{\boldsymbol B^{(h)}\}_{h \neq k}$ fixed; \\           
\hspace*{0.6cm} \textbf{until} convergence   \\
\hspace*{0.4cm} end                         \\
\hspace*{0.2cm} \textbf{until} convergence   \\ 
\hspace*{0.2cm} Compute SVD of $\boldsymbol B^{(k)}$ $= \boldsymbol U^{(k)} \boldsymbol D^{(k)} {\boldsymbol V^{(k)}}^T$, $\forall k$. The core tensor of the Tucker model is  \\
\hspace*{0.2cm} $[\![\boldsymbol D^{(1)}  {\boldsymbol V^{(1)}}^T, \boldsymbol D^{(2)}  {\boldsymbol V^{(2)}}^T,\boldsymbol D^{(3)}  {\boldsymbol V^{(3)}}^T]\!]$ and the factor matrices are $\boldsymbol U^{(k)}$, $\forall k$. \\  \hline
\end{tabular}
}
\label{tab:my-table}
\end{table}


Notice that
parameter $L$ is related to how flexible the core tensor is, and has nothing to do with the multilinear rank of the Tucker model. 
It should be noticed that with the proposed optimization formulation imposing low-rankness on $\{\boldsymbol B^{(k)}\}$, 
the multilinear rank is revealed by the number of non-zero singular values (or singular values above certain threshold if the data is noisy) from
$\boldsymbol D^{(1)}$, $\boldsymbol D^{(2)}$, $\boldsymbol D^{(3)}$ in SVD of 
$\boldsymbol B^{(1)}$, $\boldsymbol B^{(2)}$, $\boldsymbol B^{(3)}$.
In this way, the multilinear rank could be automatically learned without updating $L$.


Finally, we reveal the convergence of Algorithm 1 by establishing two properties. As Algorithm 1 is based on the BCD framework, and the optimal point for each subproblem with respect to $\boldsymbol B^{(k)}$ is obtained using (\ref{eq:20111}), the value of the objective function (\ref{eq:13}) monotonically decreases after each BCD iteration. Combining with the fact that the objective function value is bounded below, the following lemma holds.

\vbox{}
\noindent
\textbf{LEMMA 1}. Algorithm 1 guarantees the objective function value of (\ref{eq:13}) converges.

\vbox{}
Notice that the optimization problem (\ref{eq:13}) is non-convex and components $\boldsymbol B^{(k)}$ are non-linearly coupled in the
objective function. This makes the quality of convergence point under alternative minimization unknown. To find out the quality of solution of the proposed algorithm, we present the following proposition.

\vbox{}
\noindent
\textbf{PROPOSITION 1}. Algorithm 1 admits a unique limit point, and this limit point is at least a Karush-Kuhn-Tucker (KKT) point of (\ref{eq:13}).

\vbox{}
\noindent
Proof: See Appendix D.

\section{Experimental Results}  

We evaluated the proposed optimization model and method by extensive experiments and compared it with the state-of-the-art Tucker completion algorithms, including SiLRTC \cite{LRTC3}, HaLRTC \cite{LRTC3}, TREL1+W-Tucker \cite{WTucker,QShi}, CTNM \cite{CTNM}, PLM \cite{NRATC}, NNCP \cite{YLFactorMatrix}, LRTV \cite{LRTV}, McAlm \cite{McAlm} and KBR \cite{KBR}. The implementation of these competing algorithms are based on Matlab and are provided by the corresponding references. In all the experiments, the parameters of the compared methods follow the default setting suggested in their respective works. Furthermore, as HaLRTC is a noiseless model, we extend it to handle noise by replacing the projection operation with the least squares criterion at the objective function (details on the model and corresponding algorithm can be found in Appendix E), and we apply this noisy version HaLRTC whenever the data contains noise. For the proposed algorithm LRFMTC\footnote{The implementation code of the proposed LRFMTC is available at https://github.com/XKTONG.}, we set $L=150$, $\alpha = 30$ for the synthetic data and $\alpha = 5$ for image data, chemometrics data and the HSI data (Analysis on how different choices of $L$ and $\alpha$ affecting the performance could be found in Appendix F). The multilinear rank is determined by retaining singular values in $\boldsymbol D^{(k)}$ if its squared value is larger than 0.0001 times of the squared value of the largest singular values. To assess the performance, relative squared error RSE$= ||\mathcal X - \widehat{\mathcal X}||_{F}/||\mathcal X||_{F}$ and the estimated multilinear rank (if applicable) are computed, where $\widehat{\mathcal X}$ is the reconstructed tensor from various algorithms. All results in this section are averaged over 50 trials with each trial having independent missing pattern and noise realization.

\begin{figure*}[t]
\centering
\subfigure[]{
\begin{minipage}[t]{0.45\textwidth}
\centering
\includegraphics[width=7cm]{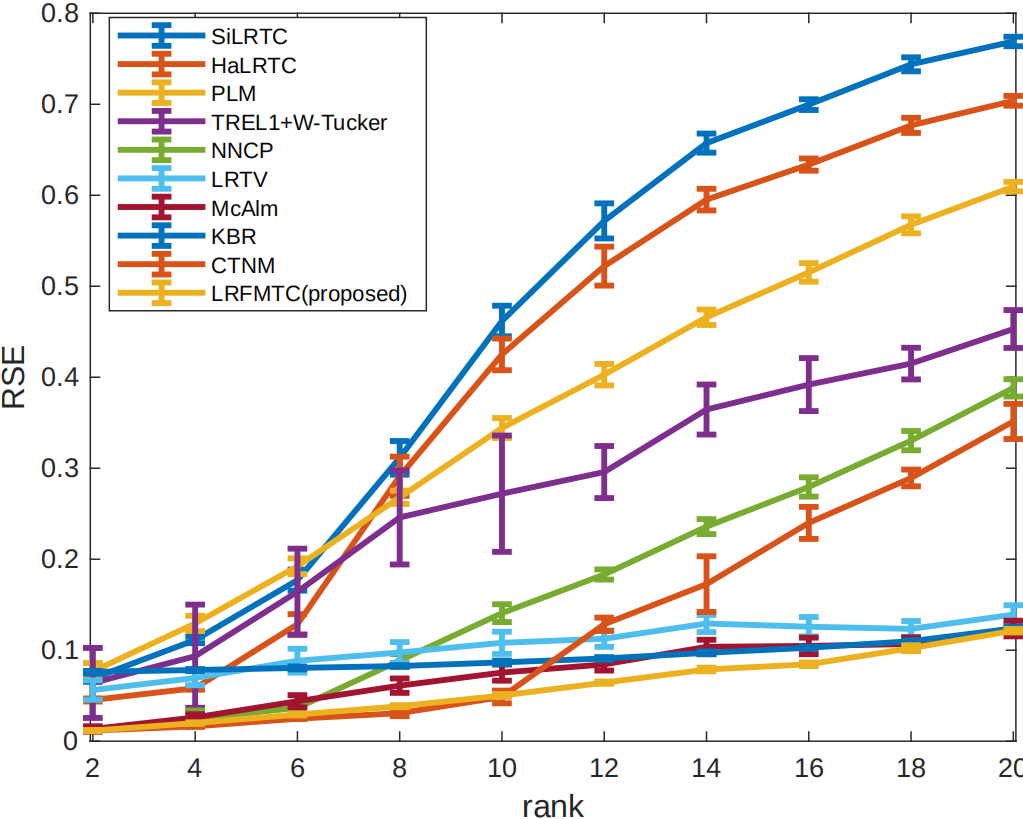}
\end{minipage}
}
\subfigure[]{
\begin{minipage}[t]{0.45\textwidth}
\centering
\includegraphics[width=7cm]{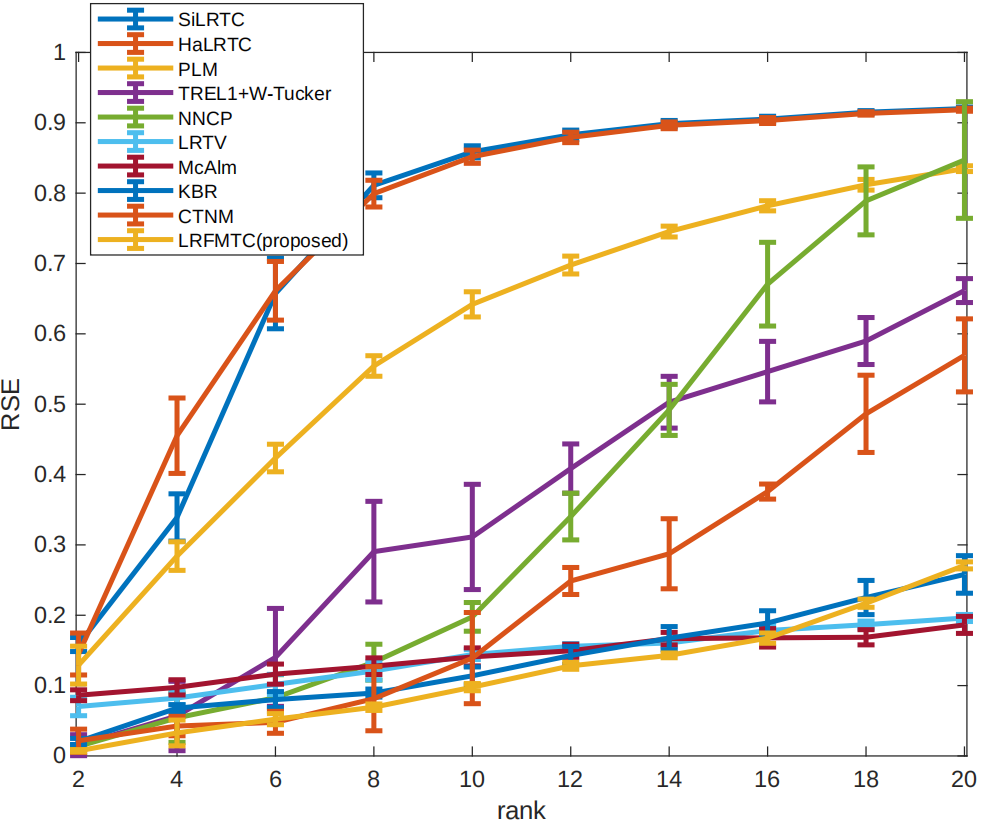}
\end{minipage}
}
\centering
\captionsetup{font={footnotesize,stretch=1}}
\caption{Averaged RSE comparison under a) sampling ratio (SR)=20\% at SNR=20dB and b) sampling ratio (SR)=10\% at no noise. The vertical error bars show one standard deviation.}
\label{RSEComparison}
\end{figure*}

\subsection{Synthetic data}
We use the synthetic Tucker data with size $\{ {I_k = 50} \}^3_{k=1} $ and the multilinear rank $\{ R_k = R \}^3_{k=1} $ where $R$ takes values 2, 4, 6, 8, · · · , 20. The elements of the core tensor and the factor matrices are i.i.d. standard Gaussian distributed, and the columns of factor matrices are orthogonalized. Two cases are considered. The first case corresponds to noisy data with Gaussian noise added to the generated synthetic tensor data at signal-to-noise ratio (SNR) \footnote{SNR= $10 \log_{10}\frac{E_S}{E_N} $, where $E_S$ is the variance of the signal and $E_N$ is the variance of random noise.} being 20dB, and the sampling ratios (SR, or percent of observations) are set to be 20$\%$. The second case is with SR=10\% but no noise is added to the synthetic data.

\begin{table*}[t] 
\begin{minipage}[t]{1\textwidth}
\centering
\captionsetup{font={footnotesize,stretch=1}}
\caption{Averaged rank estimation for the setting of Fig. \ref{RSEComparison}(a)}
\scalebox{0.5}{
\begin{tabular}{ m{5em}<{\centering} m{7em}<{\centering}  m{7em}<{\centering}   m{7em}<{\centering}  m{7em}<{\centering} m{7em}<{\centering} m{7em}<{\centering}  m{7em}<{\centering} m{7em}<{\centering} m{7em}<{\centering}  }
\hline
SR=20\%  & SiLRTC & {HaLRTC}   &CTNM & NNCP& TREL1+WTucker & McAlm& LRTV & PLM &LRFMTC \\      \hline
  True Rank      &   \multicolumn{7}{c}{Estimated Rank}     \\  
  (2 2 2)        & (2.0 2.0 2.0)   & (2.0 2.0 2.0)   &(2.0 2.0 2.0) &(2.6 2.0 2.2)&(1.9 2.0 2.2)  &  (3.0 2.9 3.0)  &  (2.6 2.8 2.3)     & (2.4 25.3 22.3) & (2.0 2.0 2.0)  \\
  (4 4 4)        & (4.1 4.3 4.2)   &(4.1 4.4 4.3)   &  (4.0 4.0 4.0)&(4.5 4.7 4.6)&(4.4 4.3 4.0)  & (4.7 4.9 4.9)    &  (5.3 5.2 5.5)   &(8.7 49.9 49.5)&(4.0 4.0 4.0)   \\ 
  (6 6 6)        & (18.4 17.6 18.5)&(27.8 26.4 27.5)&(7.0 7.0 7.0)& (7.5 7.2 7.4)   &(6.3 6.7 6.4)    &  (5.8 6.6 6.6)     &  (11.9 12.4 12.2)     &(15.5 50.0 50.0)&(6.0 6.0 6.0) \\ 
  (8 8 8)        & (48.3 47.9 47.4) &(50.0 50.0 49.9)&(8.8 8.8 8.8)&  (9.7 9.4 9.9) &(8.5 8.4 8.6) & (7.5 8.8 9.1)  & (14.3 14.4 14.8)   &(19.3 50.0 50.0)&(8.0 8.0 8.0)    \\ 
  (10 10 10)    & (50.0 50.0 50.0) &(50.0 50.0 50.0)&(10.0 10.0 10.0)& (12.9 13.3 12.4)  &(13.4 13.1 12.9) &(8.7 12.7 12.7)  & (20.8 20.7 21.0) &(22.1 50.0 50.0)&(10.0 10.0 10.0)  \\   
  (12 12 12)     &(50.0 50.0 50.0)  &(50.0 50.0 50.0)&(11.0 11.0 11.0)&  (16.2 16.0 16.3)   &(15.6 16.0 16.0)& (9.5 15.6 15.3)  &  (24.8 24.5 24.8)  &(23.1 50.0 50.0)&(12.0 12.0 12.0)   \\ 
  (14 14 14)     & (50.0 50.0 50.0) &(50.0 50.0 50.0)&(12.3 12.3 12.3)&  (20.4 20.0 19.1)   &(16.2 16.4 16.2)& (11.3 24.5 24.8)   &  (26.0 26.5 25.8)    &(22.9 50.0 50.0)&(14.0 14.0 14.0)   \\ 
  (16 16 16)     & (50.0 50.0 50.0)&(50.0 50.0 50.0)&(12.9 12.9 12.9)& (25.0 24.8 24.1)  &(18.8 18.9 18.6)  & (12.1 24.9 25.0)   &  (27.1 26.7 26.8)    &(23.4 50.0 50.0)&(16.0 16.0 16.1)   \\  
  (18 18 18)     & (50.0 50.0 50.0)&(50.0 50.0 50.0)&(14.0 14.0 14.0)& (29.7 29.0 29.0)    &(21.8 22.0 22.0) & (12.8 25.1 25.0)  &(28.2 28.4 28.4)    &(23.6 50.0 50.0)&(18.2 18.0 18.3)  \\  
  (20 20 20)   & (50.0 50.0 50.0)&(50.0 50.0 50.0)&(14.6 14.6 14.6)&  (33.9 32.9 33.3)   &(24.1 24.5 23.9) & (14.7 34.1 34.0)  &  (30.2 30.0 29.4)  &(22.8 50.0 50.0)&(21.3 21.0 22.1)  \\      \hline                                                                                                                                      
\end{tabular}
}
\label{Rank1}
\end{minipage}
\end{table*}

Figure \ref{RSEComparison} shows the average RSE and the one standard deviation (illustrated by the vertical error bar) of various Tucker completion methods. In general, we could see that NNCP and CTNM algorithms have better performance than SiLRTC, HaLRTC, PLM, and TREL1+W-Tucker. But the proposed LRFMTC algorithm performs the best, achieving the smallest RSE for most of the cases. Furthermore, the width of one standard deviation shows that the proposed LRFMTC algorithm has the smallest variation around the mean RSE compared to other algorithms. This demonstrates the performance of the proposed method is more stable across different trials than other algorithms.


\begin{table*}[t] 
\begin{minipage}[t]{1\textwidth}
\centering
\captionsetup{font={footnotesize,stretch=1}}
\caption{Averaged rank estimation for the setting of Fig. \ref{RSEComparison}(b)}
\makeatletter\def\@captype{table}
\scalebox{0.5}{
\begin{tabular}{ m{5em}<{\centering} m{7em}<{\centering}  m{7em}<{\centering}   m{7em}<{\centering}  m{7em}<{\centering} m{7em}<{\centering} m{7em}<{\centering}  m{7em}<{\centering} m{7em}<{\centering} m{7em}<{\centering}  }
 \hline
SR=10\%  & SiLRTC & HaLRTC   &CTNM & NNCP& TREL1+WTucker & McAlm& LRTV & PLM &LRFMTC \\     \hline
  True Rank      &   \multicolumn{7}{c}{Estimated Rank}     \\    
  (2 2 2)        & (2.2 2.2 2.2)   & (3.9 3.8 3.6)  &(2.0 2.0 2.0) & (2.0 2.0 2.0) & (2.2 2.0 2.5)&(2.9 14.6 14.3)&(4.4 3.9 5.2)& (2.0 21.1 20.5) & (2.0 2.4 2.0)  \\
  (4 4 4)        & (24.2 24.6 25.0) &(40.7 41.9 41.4) &(4.0 4.0 4.0)& (4.6 4.7 4.6) &(5.6 6.0 5.9)  &(4.2 16.4 16.5)&(6.4 6.7 7.4)        &(4.0 49.1 48.4)&(4.0 4.0 4.0)   \\ 
  (6 6 6)        & (49.7 49.7 49.8)&(49.9 49.9 50)&(6.0 6.0 6.0)& (7.0 7.7 7.9) & (7.2 7.3 6.8)   &(5.2 21.7 21.6)&(13.5 13.1 13.5)          &(6.7 50.0 50.0)&(6.0 6.0 6.0) \\ 
  (8 8 8)        & (50.0 50.0 50.0) &(50.0 50.0 50.0)&(8.0 8.0 8.0)&(13.3 13.4 14.2) &(9.2 9.1 9.2) &(6.4 25.0 24.8)&(15.4 16.1 15.7)           &(11.6 50.0 50.0)&(8.0 8.0 8.0)    \\  
  (10 10 10)    & (50.0 50.0 50.0) &(50.0 50.0 50.0)&(9.8 9.8 9.8)& (20.1 20.4 19.4)&(13.6 13.4 13.2)  &(8.0 29.6 29.0)&(21.0 21.2 17.7)            &(14.2 50.0 50.0)&(10.0 10.0 10.0)  \\   
  (12 12 12)     &(50.0 50.0 50.0)  &(50.0 50.0 50.0)&(10.0 10.0 10.0)&(23.7 22.9 23.6)&(14.1 14.2 14.1)  &(11.8 31.5 31.2)&(26.2 25.8 21.0)      &(14.9 50.0 50.0)&(12.0 12.0 12.0)   \\ 
  (14 14 14)     & (50.0 50.0 50.0) &(50.0 50.0 50.0)&(11.8 11.8 11.8)& (24.7 24.6 24.6)&(15.1 15.1 15.1)   &(13.5 37.7 37.3)&(28.7 28.5 27.6)       &(16.1 50.0 50.0)&(14.0,14.0,14.0)   \\ 
  (16 16 16)     & (50.0 50.0 50.0)&(50.0 50.0 50.0)&(12.0 12.0 12.0)&(30.0 30.0 30.0) &(17.9 17.3 16.9)   &(15.7 36.4 36.7)&(29.0 28.9 27.2)      &(16.9 50.0 50.0)&(16.0 16.0 16.0)   \\  
  (18 18 18)     & (50.0 50.0 50.0)&(50.0 50.0 50.0)&(12.8 12.8 12.8)& (30.0 30.0 30.0)&(21.1 21.1 20.9)  &(17.0 35.7 36.7)&(28.2 27.5 26.7)         &(17.4 50.0 50.0)&(18.8 18.6 18.7)  \\  
 (20 20 20)   & (50.0 50.0 50.0)&(50.0 50.0 50.0)&(14.2 14.2 14.2)& (30.0 30.0 30.0)&(23.4 23.5 23.8)    &(19.6 41.0 41.4)&(33.2 32.6 32.7)      &(17.7 23.5 23.8)&(22.0 22.0 22.1)  \\      \hline                                                                                                                                        
\end{tabular}
}
\label{Rank2}
\end{minipage}
\end{table*}

\begin{table*}[t]
\begin{minipage}[t]{1\textwidth}
\centering
\captionsetup{font={footnotesize,stretch=1}}
\caption{Average run time of various algorithms on synthetic data. The number on the left and right side of / corresponds to the average run time for the setting in Fig. \ref{RSEComparison}(a) and Fig. \ref{RSEComparison}(b), respectively. 
}
\makeatletter\def\@captype{table}
\scalebox{0.51}{
\begin{tabular}{ m{5em}<{\centering} m{6em}<{\centering}  m{6em}<{\centering}  m{6em}<{\centering}  m{6em}<{\centering}  m{6em}<{\centering} m{6em}<{\centering} m{6em}<{\centering} m{6em}<{\centering} m{6em}<{\centering} m{6em}<{\centering} }
\hline  
SR=20\%/10\%  & SiLRTC & HaLRTC   &CTNM & NNCP& TREL1+WTucker &McAlm &LRTV& {KBR} & PLM &LRFMTC \\     \hline
  True Rank      &   \multicolumn{5}{c}{}     \\    
  (2 2 2)        & 14.14 / 16.12  &  {5.59} / 5.35 &   1.75 / 2.15 &2.34 / 2.03 &8.70 / 7.01&2.55 / 2.10&3.82 / 3.48 &{10.19 / 9.92}& 20.66 / 27.93&13.69 / 15.69 \\
  (4 4 4)       &  14.61 / 16.33 & {5.63} / 5.42 &1.87 / 2.38 &3.75 / 3.77&19.81 / 15.19&2.49 / 2.12&3.87 / 3.50&{9.42 / 10.95}&20.87 / 25.97 & 12.81 / 15.22\\ 
  (6 6 6)        & 15.25 / 13.96 & {5.42} / 5.42 &3.48 / 2.36&7.02 / 6.40&27.69 / 22.03&2.43 / 2.11&3.93 / 3.47&{9.36 / 9.08}& 30.31 / 27.29 &14.25 / 15.40\\ 
  (8 8 8)        &  15.14 / 14.63& {5.87} / 5.60 & 3.53 / 3.07&8.66 / 8.48&30.05 / 27.50&2.43 / 2.12&3.87 / 3.55   &{9.57 / 8.40}& 34.34 / 28.72 & 16.09 / 16.44  \\ 
  (10 10 10)    &  15.18 / 15.12&  {5.91} / 5.50&3.62 / 3.41&10.54 / 10.59&34.49 / 31.73&2.47 / 2.13&3.91 / 3.45   &{11.20 / 8.18}& 35.07 / 25.02 & 19.78 / 19.39\\                                                    
  (12 12 12)    &  15.90 / 15.61& { 5.36} / 5.42&3.84 / 3.43&13.05 / 10.99&38.10 / 33.89&2.50 / 2.13&3.83 / 3.50     &{10.94 / 8.27}&36.76 / 25.27 & 20.38 / 20.57\\
  (14 14 14)     &  15.35 / 15.14& {5.20} / 5.38 & 4.27 / 3.86&14.94 / 11.42&28.39 / 36.46&2.50 / 2.16&3.91 / 3.52     &{9.78 / 8.98}& 35.64 / 23.28 &  20.75 / 22.34\\ 
  (16 16 16)     &  15.24 / 14.88& {5.58} / 5.38 &4.24 / 3.58&16.85 / 13.62&15.39 / 38.92&2.48 / 2.13&3.84 / 3.49     &{9.51 / 8.46}&37.87 / 24.57 & 22.26 / 23.13\\  
  (18 18 18)     &  15.52 / 14.75 & {5.95} / 5.45 &4.30 / 3.74&19.12 / 10.45&13.16 / 43.06&2.49 / 2.20&3.82 / 3.47    &{8.90 / 9.19}& 38.08 / 26.55 & 23.93 / 23.55\\  
  (20 20 20)  &  15.30 / 14.17 & {5.38} / 5.44& 4.40 / 4.08&21.21 / 10.62& 9.00 / 38.71&2.55 / 2.14&3.87 / 3.46    & {8.32 / 9.08}& 38.14 / 25.16 & 22.75 / 24.22\\      \hline                                                                                                                          
\end{tabular}}
\label{runtime}
\end{minipage}
\end{table*}

On the other hand, the averaged estimated multilinear rank from various algorithms are shown in Table \ref{Rank1} and Table \ref{Rank2} for SR = 20\% and 10\% respectively. It can be seen that while the state of the arts algorithms could estimate the multilinear rank correctly when the true rank is small (e.g., 2-4), their estimates completely fail when the true rank is larger than 6. Among the competing algorithms, CTMN performs relatively well. However, the proposed LRFMTC algorithm gives the closest multilinear rank estimates for the whole range of true rank. This superior performance in multilinear estimation is a major reason for the RSE of the proposed LRFMTC being much smaller than competing methods. In particular, when comparing with the NNCP \cite{YLFactorMatrix}, which starts with the same problem formulation (14) as in the proposed LRFMTC but introduces auxiliary variables in its algorithm, the proposed LRFMTC estimates the multilinear rank more accurately as there is no auxiliary variables in the proposed algorithm. This clearly shows that introducing auxiliary variables would suffer from ineffective low-rankness propagation. Table \ref{runtime} compares the average run time of various algorithms for the synthetic data. It can be seen that the proposed algorithm has a run time comparable to that of PLM, and they are slower than other algorithms.

\subsection{Image data}

Next, we show the results of image completion on 8 benchmark color images\footnote{Datasets are from https://ieeexplore.ieee.org/abstract/document/7010937/ media\#media.} each with size 256$\times$256$\times$3.
For NNCP, the initialization of CP-rank is 40 \cite{YLFactorMatrix},
and for CTNM, the upper bound of the multilinear rank is set at $(40,40,3)$ \cite{CTNM}. 
On the other hand, SiLRTC/HaLRTC does not make use of the multilinear rank information. For the results in this section, we use PSNR and SSIM as assessment
criteria.

\begin{table*}[b]
\caption{PSNR and SSIM comparison in image completion (SR=0.3) with 20dB noise}
\footnotesize
\centering
\scalebox{0.73}{
\begin{tabular}{ l |m{2em} m{2em} |m{2em} m{2em} |m{2em} m{2em} |m{2em} m{2em}|m{2em} m{2em} |m{2em} m{2em} |m{2em} m{2em} |m{2em} m{2em}|m{2em} m{2em} |m{2em} m{2em} }
 \hline
 \quad 
 & \multicolumn{2}{c|}{SiLRTC} 
 & \multicolumn{2}{c|}{{HaLRTC}} 
 & \multicolumn{2}{c|}{CTNM} 
 & \multicolumn{2}{c|}{NNCP} 
 & \multicolumn{2}{c|}{\begin{tabular}{@{}c@{}} TREL1+ \\ WTucker\end{tabular}} 
 & \multicolumn{2}{c|}{McAlm}            
 & \multicolumn{2}{c|}{LRTV}
  & \multicolumn{2}{c|}{{KBR}}
  & \multicolumn{2}{c|}{PLM} 
 & \multicolumn{2}{c}{\begin{tabular}{@{}c@{}} (Proposed) \\ LRFMTC\end{tabular}}\\
\textbf{} & PSNR & SSIM  &  PSNR &  SSIM &  PSNR &  SSIM &  PSNR &  SSIM & PSNR & SSIM & PSNR & SSIM&  PSNR &  SSIM  &  PSNR &  SSIM & PSNR & SSIM  & PSNR & SSIM \\
\hline
peppers & 23.57  &0.823&{24.09} &{0.829}&   24.31   &0.806&   23.44   &0.778&   22.14   &0.732                &22.07&0.770     &20.80&0.779 &{23.05}&{0.764}  &23.72   &0.800  & \bf{25.20}    & \bf{0.841} \\
boats    &  23.30 &0.781&  {23.51}  &{0.787}&   23.27   &0.753&  23.54    &0.774&   22.81   &0.734                 &22.31&0.731    &22.67&0.804 &{23.89}&{0.764}   &23.03   &0.752 & \bf{24.95}   &\bf{0.819}  \\
barbara & 24.68  &0.860& {24.97}   &{0.869}&  25.11    &0.864&   25.04   &0.866&  23.57    &0.815                  &23.74&0.832      &22.47&0.868  &{24.53}&{0.854}  &24.74   &0.852 & \bf{26.15}   & \bf{0.888} \\
house   &  26.49 &0.791& {26.85}  &{0.793}&    26.96  &0.763&   27.31   &0.757&   24.15   &0.640                   &24.69&0.717        &24.48&0.755  &{25.72}& {0.701} &25.74     &0.736 & \bf{27.54}   &  \bf{0.799}\\
airplane & 24.70  &0.823& { 25.33}  &{0.823}&  24.71    &0.779&   25.29   &0.795&  23.23    &0.686                 &21.22&0.743       &22.51&\bf{0.864} &{24.83}& {0.746} &24.52  &0.774  &\bf{25.34}   & 0.811 \\
sailboat & 22.56  &0.792&  {23.12}  &{0.797}&   22.34   &0.753&   22.63   &0.766&  21.48   &0.712                  &21.73&0.738           &22.28&0.790 &{22.41} &{0.743}  &22.34  &0.751   &\bf{23.67}   & \bf{0.803} \\
facade  & 28.80  &0.943&{\bf{29.11}} &{{0.946}}& 28.52 &0.943 &28.54&0.944& 27.78   &0.932                  &25.88&0.932     &21.44&0.832  &{28.28}& {\bf 0.947}  &   28.50 & 0.940  &28.80   & 0.945 \\
baboon &  21.75 &0.696&  {21.87}  &{\bf0.706}&  20.43    &0.626&   20.19   &0.633&  19.76    &0.579                   &21.00&0.669             &18.22&0.671 &{20.65}& {0.671}  &21.14  &0.671   &\bf{21.91}   &  \bf{0.706}\\
\hline
\end{tabular}}
\label{ImagePSNRSSIM1}
\end{table*}

\begin{table*}[b]
\caption{PSNR and SSIM comparison in image completion (SR=0.2) without noise.}
\footnotesize
\centering
\scalebox{0.73}{
\begin{tabular}{ l |m{2em} m{2em} |m{2em} m{2em} |m{2em} m{2em}|m{2em} m{2em} |m{2em} m{2em} |m{2em} m{2em} |m{2em} m{2em} |m{2em} m{2em} |m{2em} m{2em} |m{2em} m{2em} }
 \hline
 \quad & \multicolumn{2}{c|}{SiLRTC} 
 & \multicolumn{2}{c|}{HaLRTC} 
 & \multicolumn{2}{c|}{CTNM} 
 & \multicolumn{2}{c|}{NNCP} 
 & \multicolumn{2}{c|}{\begin{tabular}{@{}c@{}} TREL1+ \\ WTucker\end{tabular}}  
  & \multicolumn{2}{c|}{McAlm}
   & \multicolumn{2}{c|}{LRTV}
     & \multicolumn{2}{c|}{{KBR}}
    & \multicolumn{2}{c|}{PLM}
 & \multicolumn{2}{c}{\begin{tabular}{@{}c@{}} (Proposed) \\ LRFMTC\end{tabular}}\\
\textbf{} & PSNR & SSIM  & PSNR & SSIM & PSNR & SSIM &  PSNR &  SSIM  & PSNR & SSIM &  PSNR &  SSIM  &  PSNR &  SSIM & PSNR & SSIM  & PSNR & SSIM & PSNR & SSIM \\
\hline
peppers & 21.41  &  0.751 & 21.25   & 0.744 & 21.90   &  0.731&  20.80  &  0.682  & 20.37& 0.657                  &20.49&0.700      &16.91&0.685  &{22.46}&{0.762}   & 22.05 &  0.752 & {\bf 22.99} & {\bf{0.772}}  \\
boats    & 21.53  & 0.703  & 21.64  & 0.702  & 21.02  & 0.645 & 20.89 &  0.673  & 20.80  & 0.652                 &20.60&0.651      &19.83&0.705  &{22.73}&{ \bf 0.740}  & 21.46  & 0.688 & {\textbf{22.78}}  &  {0.732}        \\
barbara &  22.61 & 0.787   & 23.03 &  0.803   & 22.98  &0.790  & 22.83  &  0.800   & 21.45  &  0.721            &21.76&0.753      &19.60&0.793 &{23.38}&{0.828}  & 22.98&  0.794  & {\textbf{24.33}}  &  {\textbf{0.840}}   \\
house   & 24.39   &  0.745  & 24.15&  0.759 &24.16&0.707&24.13  &0.681  & 22.18  &  0.567             &23.20&0.673     &20.40&0.660  &{25.41}&{0.723} & 24.23 &  0.706  & {\textbf{25.95}}  & {\bf 0.783}   \\
airplane & 22.63 &  0.748 & 22.92&  0.761   & 22.18  & 0.681 & 22.54  & 0.695   & 21.18  &  0.611                  &20.30&0.670     &21.91&\textbf{0.842} &{23.37}&{ 0.734}& 22.71  &  0.721 &{\textbf{24.15}} &  {0.756}  \\
sailboat & 20.83 &  0.722 &  21.06 &  0.726& 20.32    &  0.665 & 20.58  &  0.685   & 19.15  &   0.616            &20.14&0.663      &19.41&0.698    &{21.54}&{ 0.719}    & 20.82 & 0.691   & {\textbf{22.34}}& {\textbf{0.741}}  \\
facade  & 27.21  &0.916  &27.68 &0.924 &27.13 &0.919 &26.46 &0.910&26.65 &0.911             &24.87&0.907    &18.70&0.698      &{\bf 27.75}& {\bf 0.938}  & 27.33 &0.920    &{26.85} &  {0.911}  \\
baboon &20.46  &   0.596  & 20.58 &0.599  &19.43  &0.545  &18.56  &0.515  &18.32 &0.478           &19.72&0.567      &15.57&0.536   &{19.94}& {0.603}   &  19.95& 0.574   &{\textbf{20.72}}   & { \bf 0.605}   \\
\hline
\end{tabular}}
\label{ImagePSNRSSIM2}
\end{table*}



\begin{table*}[t] 
\begin{minipage}[t]{1\textwidth}
\centering
\captionsetup{font={footnotesize,stretch=1}}
\caption{Average run time of various algorithms on image data. The number on the left and right side of / corresponds to the average run time for the setting in Table \ref{ImagePSNRSSIM1} and Table \ref{ImagePSNRSSIM2}, respectively.}
\makeatletter\def\@captype{table}
\scalebox{0.5}{
\begin{tabular}{ m{5em}<{\centering} m{6em}<{\centering}  m{6em}<{\centering}  m{6em}<{\centering}  m{6em}<{\centering} m{8em}<{\centering} m{5em}<{\centering}  m{6em}<{\centering} m{6em}<{\centering} m{6em}<{\centering} m{6em}<{\centering} }
  \hline
SR=30\%/20\%
& SiLRTC 
& HaLRTC   
&CTNM 
& NNCP
& TREL1+WTucker
& McAlm 
& LRTV
& {KBR}
& PLM 
&LRFMTC \\     \hline
  Peppers      &  75.54 / 70.49& {16.47} / 16.90 & 5.92 / 5.24&31.64 / 25.32 &62.46 / 58.33&4.38 / 5.26&3.38 / 3.09&{21.91 / 21.80}&20.42 / 18.93& 26.15 / {33.87} \\
  Boats         &  80.71 / 77.80& {18.29} / 16.36 & 5.74 / 4.50&28.65 / 24.49 &70.77 / 60.66    &5.78 / 4.56 &  3.92 / 3.06  & {21.93 / 21.06}&21.15 / 18.68& 27.82 / {34.90}\\ 
  Barbara     & 82.02 / 71.64 & {17.36} / 16.45 &5.63 / 4.64&30.27 / 26.27 & 58.10 / 57.57 &4.31 / 4.22 & 3.97 / 3.05 &{23.26 / 21.47}&19.12 / 18.42& 27.60 / {35.85}\\ 
  House       & 82.16 / 75.97 & {17.08}  / 15.83 &5.50 / 4.58&29.10 / 25.66 & 72.30 / 63.78 &4.32 / 4.12 &  3.41 / 3.14 & {22.27 / 20.75}& 19.09 / 18.51&  25.57 / {32.94}\\ 
  Airplane & 65.78 / 86.34 & {16.89} / 15.91 & 5.56 / 4.57&30.48 / 24.98 &  74.58 / 59.54   &5.35 / 4.96 & 3.75 / 3.14 & {20.89 / 19.98} &18.98 / 18.20& 26.36  / {31.78}\\                   
  Sailboat & 72.70 / 79.76 & {16.92} / 16.43 & 5.53 / 4.63 &31.03 / 25.73 & 66.73 / 60.71 &5.19 / 4.90 &  3.62 / 3.14  & {22.30 / 22.48}& 19.54 / 18.02& 27.57 / {36.56}\\ 
  Facade     & 79.63 / 83.84 & {16.53} / 15.97 & 5.60 / 4.52&28.81 / 24.98 &  60.77 / 56.71  & 4.50 / 4.20 &  3.52 / 3.13 &{22.74 / 22.11}& 19.66 / 18.80& 25.21 / {33.67}\\ 
  Baboon     &  85.93 / 82.51& {16.77} / 16.14 & 5.47 / 5.14&31.05 / 25.21 & 65.92 / 60.84   & 4.66 / 4.31 & 3.43 / 3.55 &{22.08 / 22.32}&19.09 / 21.96& 24.88 / {35.97} \\      \hline                             
\end{tabular}}
\label{runtimeimage}
\end{minipage}
\end{table*}

The average PSNR and SSIM are shown in Table \ref{ImagePSNRSSIM1} and Table \ref{ImagePSNRSSIM2} for the cases of SR=30\% with SNR=20dB, and SR=20\% without noise, respectively. It can be seen that the proposed method performs the best in terms of PSNR for seven out of eight tested images (in both Tables \ref{ImagePSNRSSIM1} and \ref{ImagePSNRSSIM2}). Furthermore, for SSIM, the proposed method achieves six best and one second best results out of eight images for the case SR=0.3 and SNR=20dB. For the SR=0.2 without noise case, the proposed method achieves four best and one second best SSIM out of eight tested images. Visual differences of image completion due to different algorithms are shown in Figure \ref{SampleImages} of Appendix G.

The average run times of various algorithms on the images are shown in Table \ref{runtimeimage}. It can be seen that the proposed LRFMTC has a run time similar to PLM and NNCP, and they are faster than SiLRTC and TREL1+WTucker. Further experimental results on hyperspectral images are included in the Appendix H.

\subsection{Chemometrics data}

We also evaluated the proposed LRFMTC model in chemometrics data analysis\footnote{Datasets are available from the repository for multi-way data analysis. http://www.models.kvl.dk/datasets}. 
Two data sets are chosen: Amino Acids Fluorescence (5 $\times$ 201 $\times$ 61) and Sugar Process (268 $\times$ 571 $\times$ 7).
Amino acid fluorescence dataset consists of five simple laboratory-made samples measured by fluorescence on a PE LS50B spectrofluorometer, with each sample containing different amounts of tyrosine, tryptophan and phenylalanine dissolved in phosphate buffered water. Since each individual amino acid gives a rank-one contribution to the data, the tensor is approximately rank-(3, 3, 3).
Sugar Process dataset consists of 268 samples measured spectrofluorometrically on a PE LS50B spectrofluorometer. In the third-order tensor data, the first mode refers to samples, the second mode to emission wavelengths, and the third mode to excitation wavelengths.

\begin{figure*}[t]
\centering
\subfigure[]{
\begin{minipage}[t]{0.45\textwidth}
\centering
\includegraphics[width=7cm]{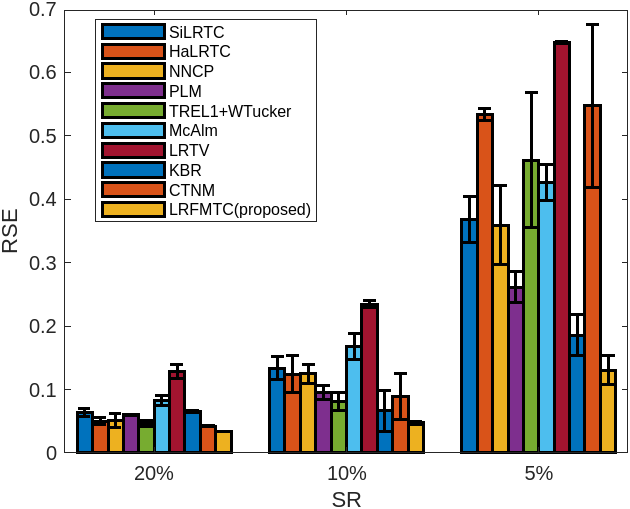}
\end{minipage}
}
\subfigure[]{
\begin{minipage}[t]{0.45\textwidth}
\centering
\includegraphics[width=7cm]{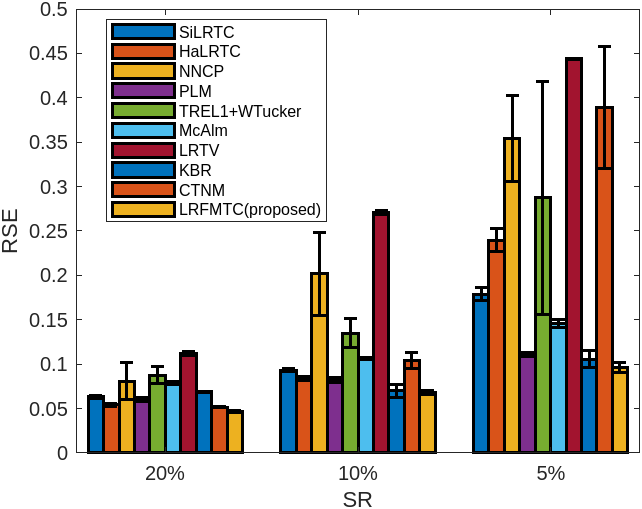}
\end{minipage}
}
\centering
\captionsetup{font={footnotesize,stretch=1}}
\caption{Averaged RSE comparison on a) Amino Acids Fluorescence and b) Sugar Process at SNR=20dB. The vertical error bars show one standard deviation of the RSE across multiple trials.}
\label{RSEAcid}
\end{figure*}

In Figure \ref{RSEAcid}(a), the average RSEs comparison show that among all the methods, the proposed LRFMTC method recovered the Amino acid fluorescence data the best under SR= 20$\%$, 10$\%$, 5$\%$. Figure \ref{RSEAcid}(b) shows the corresponding results of the Sugar Process data, and it can be observed that the proposed method could also obtain the smallest RSE, achieving the best recovery performance. Furthermore, the proposed method also attains the smallest one standard deviation among all the competing methods, showing its stable performance.

\begin{table*}[t]  
\begin{minipage}[t]{1.0\textwidth}
\centering
\captionsetup{font={footnotesize,stretch=1}}
\caption{Averaged rank estimation for the Amino acid fluorescence data}
\makeatletter\def\@captype{table}
\scalebox{0.52}{
\begin{tabular}{ m{3.3em}<{\centering} m{6.3em}<{\centering}  m{6.3em}<{\centering}  m{6.3em}<{\centering}  m{6.3em}<{\centering} m{6.3em}<{\centering} m{6.3em}<{\centering}  m{6.7em}<{\centering} m{6.3em}<{\centering} m{6.3em}<{\centering} }
  \hline
  & SiLRTC & HaLRTC   &CTNM  &NNCP & TREL1 &McAlm&LRTV& PLM  &LRFMTC \\     \hline
         &   \multicolumn{7}{c}{Estimated Rank}     \\    
  SR=20\%       & (3.6 3.1 5.0)   & (5.0 3.6 5.0)   &(6.0 5.9 4.8)& (13.0 10.0 5.0)& (5.7 12.9 5.0) &(19.9 16.1 5.0)&(109.9 57.3 5.0)& (6.0 6.0 5.0) & (3.0 3.0 3.0)  \\
  SR=10\%        & (5.6 4.0 5.0)   &(7.5 5.2 5.0)   &  (6.0 6.0 4.8)&(10.8 14.1 5.0) &(5.2 11.4 5)&(19.9 23.6 5.0)&(132.1 59.1 5.0)&(5.0 5.0 5.0)&(3.0 3.0 3.0)   \\ 
  SR=5\%       & (20.6 15.6 5.0)&(50.3 35.2 5.0)&(6.0 6.0 4.7)& (9.0 15.5 5.0)& (8 14.3 5)&(21.7 40.3 5.0)&(159.6 61.0 5.0)&(3.7 3.7 5)&(3.0 3.0 3.0) \\ 
      \hline                                        
\end{tabular}
}
\label{Rank3}
\end{minipage}
\end{table*}


\begin{table*}[b]  
\begin{minipage}[t]{1\textwidth}
\centering
\captionsetup{font={footnotesize,stretch=1}}
\caption{Average run time of various algorithms under the setting of Fig. \ref{RSEAcid}(a). 
}
\makeatletter\def\@captype{table}
\scalebox{0.63}{
\begin{tabular}{ m{4em}<{\centering} m{4em}<{\centering}  m{4em}<{\centering}  m{4em}<{\centering}  m{4em}<{\centering}  m{6.5em}<{\centering} m{5em}<{\centering} m{4em}<{\centering} m{4em}<{\centering} m{4em}<{\centering} m{4em}<{\centering} }
 \hline
  & SiLRTC & {HaLRTC}   &CTNM & NNCP& TREL1+WTucker &McAlm&LRTV&{KBR}& PLM &LRFMTC \\     \hline
    SR=20\%        & 71.51  &{5.36}   & 2.53&5.10&34.77&5.30&2.50&{10.67}&19.04 & 12.60 \\
  SR=10\%       & 73.62  &{5.21}  & 2.94&5.29&31.59&5.64&3.27&{10.06}& 56.27& 12.13 \\ 
    SR=5\%        & 70.41  &  {4.97} &2.55 &5.44&31.46&5.52&3.13&{9.42}&53.81 & 13.53\\      \hline                                                             
\end{tabular}}
\label{runtimeacid}
\end{minipage}
\end{table*}



On the other hand, the number of amino acid in Amino acid fluorescence data is known, leading to the multilinear rank-(3, 3, 3), which could be regarded as the benchmark rank. Table \ref{Rank3} gives the rank estimation of the recovered Amino acid fluorescence data from all the methods, and we could see that only the proposed LRFMTC could estimate the rank accurately.
Finally, the average run time of different algorithms for the amino acid experiments are shown in Table \ref{runtimeacid}. It can be seen that in this application, the proposed method is faster than SiLRTC, TREL1+WTucker and PLM.

\section{Conclusion}

We have revealed that existing modelings of Tucker tensor completion do not directly enforce low-rankness on the recovered tensor and thus do not provide accurate
estimate of the multilinear rank. Based on a newly discovered equivalence between the Tucker format and a CPD with factor matrices having large number of columns but being low-rank, a new formulation for Tucker completion is established to solve the long-standing challenge of multilinear rank estimation problem. From the representation capability perspective, the CPD-based new representation is the same as the standard Tucker, but the 
equivalent representation of the Tucker format gave formal justification that low-rankness should be imposed on factor matrices. This also gets rid of auxiliary variables which commonly present in existing Tucker completion algorithms, and offers direct learning of multilinear rank from tensor data.
%
%
To model the low rank structure in the factor matrices, we employed trace norm minimization on the CPD factor matrices. Then BCD and accelerated fixed point iteration have been used to derive a convergence guarantee iterative update algorithm. 
Extensive experiments on synthetic data have shown that the proposed LRFMTC algorithm 
achieves more accurate multilinear rank estimation, and smaller tensor recovery errors compared to 
state-of-the-art Tucker completion methods. Experiments on images and chemometrics application validated the superiority of the proposed method in real-world datasets. 
%

Although we used trace norm to impose low-rankness in this paper, the proposed formulation offers a new perspective on Tucker completion problem, and any low-rank regularizers can be used to replace the trace norm in the new formulation. 
%
%
While intuitive appealing and performs well empirically, this paper did not include theoretical guarantee and error bounds showing the proposed formulation (due to the new Tucker format interpretation) would lead to lower estimation error compared to other commonly used approximate problem formulations. This is a limitation of the current study and should be addressed in the future.
Furthermore, the complexity of the algorithm may be reduced by advances in optimization theory and techniques. This opens up many possibilities in developing even better Tucker completion algorithms in the future.







 


 \begin{appendices}

 \section{}
 
We prove the result for $k=1$ below, and the corresponding results for $k=2, 3$ can be obtained similarly. First, we prove two useful results before establishing the relationship between multilinear rank and $n$-rank.

(\textbf{a}) The unfolded core tensor $\boldsymbol G_{(1)}$ will be of full rank at the minimum of $\sum_{k=1}^3 R_k$. We prove this by contradiction. Notice that the dimension of $\boldsymbol G_{(1)}$ is $R_1 \times R_2R_3$. Suppose $\boldsymbol G_{(1)}$ is not full rank, then $\boldsymbol G_{(1)}$ can be factorized as a product of two matrices $\boldsymbol G_{(1)} = \boldsymbol G_1 \boldsymbol G_2$, where $\boldsymbol G_1 \in \mathbb R^{(R_1,r)}$, $\boldsymbol G_2 \in \mathbb R^{(r,R_2R_3)}$, and $r < R_1$, $r < R_2R_3$.  Replacing $\boldsymbol G_{(1)}$ with $\boldsymbol G_1 \boldsymbol G_2$ in (3), we obtain that $\boldsymbol G_1$ could be multiplied to $\boldsymbol A^{(1)} $ to form the new factor matrix with size being $(I_1,r)$ and the mode-1 unfolding of the new core becomes $\boldsymbol G_2$.  If $\boldsymbol G_2$ is folded back to the core tensor, its size is $(r,R_2,R_3)$. Therefore, if $\boldsymbol G_{(1)}$ is not of full rank, we can always construct another Tucker representation with $r+R_2+R_3 < R_1+R_2+R_3$, which contradicts with the assumption $\sum_{k=1}^3 R_k$ is at minimum.  This makes $\boldsymbol G_{(1)}$ being full rank at the minimum values of $\sum_{k=1}^3 R_k$.

(\textbf{b}) $R_1 \leq R_2R_3$ always holds at the minimum of $\sum_{k=1}^3 R_k$.  This can also be proved by contradiction. Suppose $R_1 > R_2R_3$. Using the result from part (a) that $\boldsymbol G_{(1)}$ is of full rank at the minimum values of $\sum_{k=1}^3 R_k$, it leads to the decomposition $\boldsymbol G_{(1)} = \boldsymbol G_3 \boldsymbol G_4$ where $\boldsymbol G_3 \in \mathbb R^{(R_1,R_2R_3)}$ and $\boldsymbol G_4 \in \mathbb R^{(R_2R_3,R_2R_3)}$. Replacing $\boldsymbol G_{(1)}$ with $\boldsymbol G_3 \boldsymbol G_4$ in (3), $\boldsymbol G_3$ multiplied to $\boldsymbol A^{(1)} $ forms the new factor matrix with size being $(I_1,R_2R_3)$ and the mode-1 unfolding of the new core becomes $\boldsymbol G_4$.  If $\boldsymbol G_4$ is folded back to the core tensor, its size is $(R_2R_3,R_2,R_3)$. This gives the sum of three dimensions of $\mathcal G$ being $R_2R_3+R_2+R_3$, smaller than $R_1+R_2+R_3$, which contradicts with the minimum of $\sum_{k=1}^3 R_k$ assumption at the beginning.

Next, we prove the main result: $\text{rank}(\boldsymbol X_{(1)}) = R_1$ at the minimum of $\sum_{k=1}^3 R_k$. Since $\boldsymbol A^{(k)} $ are assumed to be of full rank, we have $\text{rank}(\boldsymbol A^{(1)}) = R_1$ and $\text{rank}(\boldsymbol A^{(3)} \otimes \boldsymbol A^{(2)}) = R_2R_3$. On the other hand, by the results of part (a) and (b) above, $\text{rank}(\boldsymbol G_{(1)}) = R_1$. Therefore, $\text{rank}(\boldsymbol A^{(1)} \boldsymbol G_{(1)}) = R_1$. Furthermore, as $\boldsymbol A^{(3)} \otimes \boldsymbol A^{(2)}$ is a full-rank matrix,  $\text{rank}( \boldsymbol A^{(1)} \boldsymbol G_{(1)} (\boldsymbol A^{(3)} \otimes \boldsymbol A^{(2)} )^T) = R_1$. Since $\boldsymbol A^{(1)} \boldsymbol G_{(1)}(\boldsymbol A^{(3)} \otimes \boldsymbol A^{(2)} )^T=\boldsymbol X_{(1)}$ (by (3)), we finally have  $\text{rank}(\boldsymbol X_{(1)}) = R_1$.

\section{}

Given a trace norm minimization problem:
\begin{align}    \label{eq:16} 
\underset{\boldsymbol X} \min  \left\{ \alpha ||\boldsymbol X||_{*}  + 
     \frac{1}{2} ||  \mathcal A(\boldsymbol X) - \boldsymbol C 
                        ||^2_F      \right\},                  
\end{align}
the fixed point iteration algorithm \cite{FPC} solves it with initializing ${\boldsymbol X}^0$, 
and 
then update ${\boldsymbol X}$ by
\begin{align}      \label{eq:17} 
&\boldsymbol Z^t = \boldsymbol X^t - \tau \mathcal A^*(  \mathcal A(\boldsymbol X^t)  -  \boldsymbol C   )       \nonumber \\
&\boldsymbol X^{t+1} = D_{\tau \alpha} (\boldsymbol Z^t),
\end{align}
where $\mathcal A^*(  \mathcal A(\boldsymbol X^t)  -  \boldsymbol C   )$ is the gradient of $\frac{1}{2} ||  \mathcal A(\boldsymbol X) - \boldsymbol C  ||^2_F$ at the point $\boldsymbol X^t$, 
and 
$D_{\tau \alpha}(\boldsymbol X)$ is the singular value shrinkage operator 
\cite{FPC} defined by:
\begin{align}
D_{\tau \alpha}(\boldsymbol X):= \boldsymbol U D_{\tau \alpha}(\boldsymbol \Sigma) \boldsymbol V^T,  
\end{align}
where $\boldsymbol \Sigma = \text{diag} (\left\{\sigma_i  \right\})$ is the singular value matrix obtained by SVD of the matrix $\boldsymbol X$, and $D_{\tau \alpha}(\boldsymbol \Sigma) = \text{diag} ( \left\{\sigma_i - \tau \alpha \right\}_+)$.
Besides, in order to guarantee convergence, $\tau$ should be selected such that 
$\tau \in (0,2/\lambda_{max}({\boldsymbol A}^T\boldsymbol A))$ \cite{FPC} 
where $\boldsymbol A$ 
is a matrix satisfying
 $\text{vec}(\mathcal A(\boldsymbol X)) = \boldsymbol A \text{vec}(\boldsymbol X)$.

\vbox{}
Comparing (\ref{eq:16}) to the subproblem
(15), we have $\boldsymbol X = \boldsymbol B^{(k)}$, 
$\mathcal A(\boldsymbol X) = \mathcal A(\boldsymbol B^{(k)})= \Big[ \boldsymbol B^{(k)} \Big( \underset{h \ne k}{\bigodot}  \boldsymbol B^{(h)} \Big)^{T}  \Big] \ast  \boldsymbol O_{(k)}$ 
and $\boldsymbol C = \boldsymbol Y_{(k)} \ast \boldsymbol O_{(k)}$.  To compute the gradient of $ \frac{1}{2} || \mathcal A(\boldsymbol X) - \boldsymbol C||^2_F$, we first expand

\begin{align}     \label{eq:18} 
&\frac{1}{2} \left \| \bigg[ \boldsymbol B^{(k)} \Big( \underset{h \ne k}{\bigodot}  \boldsymbol B^{(h)} \Big)^{T} - \boldsymbol Y_{(k)}  \bigg] \ast \boldsymbol O_{(k)} \right\|^{2}_F    \nonumber \\
= & \ \frac{1}{2} \text{Tr}  \Bigg\{  
\bigg[ \bigg(
\Big( \underset{h \ne k}{\bigodot}  \boldsymbol B^{(h)} \Big) {\boldsymbol B^{(k)}}^T - {\boldsymbol Y_{(k)}}^T
\bigg)  \ast {\boldsymbol O_{(k)}}^T \bigg]    
\cdot 
\bigg[\boldsymbol O_{(k)} \ast \bigg( 
\boldsymbol B^{(k)} \Big( \underset{h \ne k}{\bigodot}  \boldsymbol B^{(h)} \Big)^{T} - \boldsymbol Y_{(k)} 
\bigg) \bigg]
 \Bigg\}  \nonumber \\
  \propto &  \  \frac{1}{2} \text{Tr} \Bigg\{
 \bigg[ \bigg(  \Big( \underset{h \ne k}{\bigodot}  \boldsymbol B^{(h)} \Big) {\boldsymbol B^{(k)}}^T  \bigg) \ast {\boldsymbol O_{(k)}}^T \bigg]       
 \cdot 
 \bigg[ \boldsymbol O_{(k)} \ast \bigg(
  \boldsymbol B^{(k)} 
 \Big(\underset{h \ne k}{\bigodot}  \boldsymbol B^{(h)} \Big)^{T}
 \bigg) \bigg]    \nonumber \\
 &\ \ \ - \bigg[ \bigg( \boldsymbol B^{(k)} \Big( \underset{h \ne k}{\bigodot}  \boldsymbol B^{(h)} \Big)^{T}  \bigg) \ast \boldsymbol O_{(k)} \bigg]    
\Big(  {\boldsymbol Y_{(k)}}^T \ast {\boldsymbol O_{(k)}}^T \Big)   \nonumber \\
 &\ \ \ 
 - \bigg[ \bigg( \Big( \underset{h \ne k}{\bigodot}  \boldsymbol B^{(h)} \Big){\boldsymbol B^{(k)}}^T   \bigg) \ast {\boldsymbol O_{(k)}}^T  \bigg] \Big( \boldsymbol Y_{(k)} \ast \boldsymbol O_{(k)} \Big)
 \Bigg\}.      
\end{align} 
Taking gradient of (\ref{eq:18}) with respect to $\boldsymbol B^{(k)}$ gives
\begin{align}      \label{eq:19} 
&\partial  \Bigg\{ \frac{1}{2}  \left \| \bigg[ \boldsymbol B^{(k)} \Big( \underset{h \ne k}{\bigodot}  \boldsymbol B^{(h)} \Big)^{T} - \boldsymbol Y_{(k)}  \bigg] \ast \boldsymbol O_{(k)} 
                        \right\|^{2}_F  \Bigg\} /\partial (\boldsymbol B^{(k)} )  \nonumber \\
 = &   \frac{1}{2}
   \Bigg\{   
2 \bigg[ \bigg( \boldsymbol B^{(k)} \Big( \underset{h \ne k}{\bigodot}  \boldsymbol B^{(h)} \Big)^{T} \bigg) \ast \boldsymbol O_{(k)} \bigg]
 \Big( \underset{h \ne k}{\bigodot}  \boldsymbol B^{(h)} \Big)    
 - 2 \Big( \boldsymbol Y_{(k)} \ast \boldsymbol O_{(k)} \Big)
\Big( \underset{h \ne k}{\bigodot}  \boldsymbol B^{(h)} \Big)
    \Bigg\}                          \nonumber \\
 = &     
\bigg[ \bigg( \boldsymbol B^{(k)} \Big( \underset{h \ne k}{\bigodot}  \boldsymbol B^{(h)} \Big)^{T}
 -  \boldsymbol Y_{(k)} \bigg) \ast \boldsymbol O_{(k)} \bigg] \Big( \underset{h \ne k}{\bigodot}  \boldsymbol B^{(h)} \Big)                       
\end{align}
Therefore $\mathcal A^*(  \mathcal A(\boldsymbol X^t)  -  \boldsymbol C   )$ in the context of subproblem (15) is
\begin{align}         \label{eq:20} 
&\mathcal A^*(  \mathcal A({\boldsymbol B^{(k)}}^t)  -  \boldsymbol Y_{(k)} \ast  \boldsymbol O_{(k)}  ) \nonumber \\
= &   \bigg[ \bigg( {\boldsymbol B^{(k)}}^t \Big( \underset{h \ne k}{\bigodot}  \boldsymbol B^{(h)} \Big)^{T}
 -  \boldsymbol Y_{(k)} \bigg) \ast \boldsymbol O_{(k)} \bigg]  \Big( \underset{h \ne k}{\bigodot}  \boldsymbol B^{(h)} \Big).
\end{align}
%
Putting (\ref{eq:20}) into (\ref{eq:17}), we obtain the update equation in (16).

To obtain a range of $\tau_k$ for convergence guarantee, we need to figure out what is the matrix $\boldsymbol A$.
Notice that $\boldsymbol A$ comes from $\boldsymbol A \text{vec}(\boldsymbol X) = \text{vec}(\mathcal A(\boldsymbol X))$, and $\text{vec}(\mathcal A(\boldsymbol X))$ in the subproblem (15) is  
\begin{align}      \label{eq:600} 
&\text{vec} \big[ \boldsymbol B^{(k)} \big( \underset{h \ne k}{\bigodot}  \boldsymbol B^{(h)} \big)^{T} \big]  \ast  \text{vec} \big( \boldsymbol O_{(k)} \big)    \nonumber \\
= & \big[ \big( \big( \underset{h \ne k}{\bigodot}  \boldsymbol B^{(h)} \big) \otimes \boldsymbol I \big)    \text{vec} \big( \boldsymbol B^{(k)} \big)  \big]    \ast  \text{vec} \big( \boldsymbol O_{(k)} \big).
\end{align}
If we denote $ \boldsymbol H = \big( \big( \underset{h \ne k}{\bigodot}  \boldsymbol B^{(h)} \big) \otimes \boldsymbol I \big)  $, 
then 
$ \boldsymbol H \text{vec} \big( \boldsymbol B^{(k)} \big) = [ \boldsymbol H_{1,:} \text{vec} \big( \boldsymbol B^{(k)} \big), \boldsymbol H_{2,:}  \text{vec} \big( \boldsymbol B^{(k)} \big),..., $ $ \boldsymbol H_{I_1I_2I_3,:} \text{vec} \big( \boldsymbol B^{(k)} \big) ]^T$. 
That is, the $i^{th}$ element of (\ref{eq:600}) is denoted 
$[ \boldsymbol H_{i,:} \text{vec} \big( \boldsymbol B^{(k)} \big) ]   [\text{vec} \big( \boldsymbol O_{(k)} \big) ]_{i}
$, where $\text{vec} \big( \boldsymbol O_{(k)} \big)$ only affects the rows of $\boldsymbol H$. Therefore we could re-express (\ref{eq:600}) as 
$ \Big[\boldsymbol H \ast \big( \boldsymbol 1^T \otimes \text{vec} \big( \boldsymbol O_{(k)} \big) \big) \Big]  \text{vec} \big( \boldsymbol B^{(k)} \big) $.  
Then we obtain $\boldsymbol A = \big[ \boldsymbol H \ast  \big( \boldsymbol 1^T \otimes \text{vec} \big( \boldsymbol O_{(k)} \big) \big)  \big]$.

We should notice that $\boldsymbol A$ is obtained from $\boldsymbol H$ with some rows of $\boldsymbol H$ 
replaced by zero row vectors. Let
the removed rows of $\boldsymbol H$ 
be denoted by 
$\boldsymbol H_0$, 
such that $\boldsymbol A + \boldsymbol H_0 = \boldsymbol H$.  Since the non-zero rows of $\boldsymbol A$ and $\boldsymbol H_0$ are at different positions, $\boldsymbol A^T \boldsymbol H_0 = \boldsymbol H_0^T \boldsymbol A= 0$ and we have 
$\boldsymbol A^T \boldsymbol A + \boldsymbol H^T_0 \boldsymbol H_0 = \boldsymbol H^T \boldsymbol H$.
As $\boldsymbol H_0^T \boldsymbol H_0$ is positive semidefinite,
$\lambda_{max}({\boldsymbol A}^T\boldsymbol A) \leq \lambda_{max}({\boldsymbol H}^T\boldsymbol H)$. 
With the definition of $\boldsymbol H$,  
\begin{align}
\lambda_{max}(\boldsymbol H^T \boldsymbol H) = \ &\lambda_{max} {\Big( 
\big( \big( \underset{h \ne k}{\bigodot}  \boldsymbol B^{(h)} \big) \otimes \boldsymbol I \big)^T
\big( \big( \underset{h \ne k}{\bigodot}  \boldsymbol B^{(h)} \big) \otimes \boldsymbol I \big) \Big)}   \nonumber \\
= \ &\lambda_{max} \Big( \big[ \big( \underset{h \ne k}{\bigodot} \boldsymbol B^{(h)}  \big)^T \big( \underset{h \ne k}{\bigodot}  \boldsymbol B^{(h)} \big) \big] \otimes \boldsymbol I_{I_k} \Big)            \nonumber \\
= \ &\lambda_{max}  \Big( \big( \underset{h \ne k}{\bigodot} \boldsymbol B^{(h)}  \big)^T \big( \underset{h \ne k}{\bigodot}  \boldsymbol B^{(h)} \big) \Big) \cdot \lambda_{max}(\boldsymbol I_{I_k})         \nonumber \\
= \ &\lambda_{max}  \Big( \big( \underset{h \ne k}{\bigodot} \boldsymbol B^{(h)}  \big)^T \big( \underset{h \ne k}{\bigodot}  \boldsymbol B^{(h)} \big) \Big).
\end{align} 
Based on the characteristics of Khatri–Rao product, $\big( \underset{h \ne k}{\bigodot}  \boldsymbol B^{(h)} \big)^T \big( \underset{h \ne k}{\bigodot}  \boldsymbol B^{(h)} \big)
= {\oast}_{h \ne k}   ({\boldsymbol B^{(h)}}^T \boldsymbol B^{(h)} )
$
%
where
$ {\oast}_{h \ne k}  ( {\boldsymbol B^{(h)}}^T \boldsymbol B^{(h)} ) = ( {\boldsymbol B^{(3)}}^T \boldsymbol B^{(3)} ) \ast \cdot \cdot \cdot \ast  ( {\boldsymbol B^{(k+1)}}^T \boldsymbol B^{(k+1)} ) \ast  ( {\boldsymbol B^{(k-1)}}^T \boldsymbol B^{(k-1)} ) \ast \cdot \cdot \cdot  \ast    ( {\boldsymbol B^{(1)}}^T \boldsymbol B^{(1)} )$.
Therefore,
$\lambda_{max}({\boldsymbol H}^T\boldsymbol H) = \lambda_{max} \big({\oast}_{h \ne k}  ( {\boldsymbol B^{(h)}}^T$ $ \cdot \boldsymbol B^{(h)} )  \big)$, 
which in turns implies $\lambda_{max}({\boldsymbol A}^T\boldsymbol A) \leq \lambda_{max} \big({\oast}_{h \ne k}  ( {\boldsymbol B^{(h)}}^T \boldsymbol B^{(h)} )  \big) $
where the equality holds when elements of $\mathcal O$ are all one. Based on the inequality above, $\tau_k \in \Big(0,2/\lambda_{max} \big( {\oast}_{h \ne k}   ({\boldsymbol B^{(h)}}^T \boldsymbol B^{(h)} ) \big) \Big)$ could make sure the convergence of the algorithm.

\section{}

For initialization, we cannot follow the usual practice of zero initialization in low-rank matrix optimization, as setting $\boldsymbol B^{(k)}=0$ for all $k$ would lead to $\boldsymbol B^{(k)}$ stuck at zero matrix in all iterations. Fortunately, since $\{\boldsymbol B^{(k)} \in \mathbb R ^{I_{k} \times L} \}$ are factor matrices (with a large number of columns) of the CPD of $\mathcal Y$, we could initialize them by the CPD of $\mathcal Y$ with a large but fixed $L$. Since $L$ is fixed, the CPD can be performed using the standard alternating least squares (ALS) implementation \cite{NDSidiropoulosTDSPML}. If the observed data tensor has missing elements, we can fill the missing values by the mean of the observed data.

On the other hand, the stopping criterion for the inner iteration is when the normalized root mean square error between the estimated $\boldsymbol B^{(k)}$ of two adjacent inner iterations is smaller than $10^{-4}$ or reaches a pre-defined maximum inner iteration number.  For outer iteration, the stopping criterion is when the normalized root mean square error between the estimated tensors of two adjacent outer iterations is smaller than $10^{-6}$ or when certain maximum outer iteration number is reached.

For the proposed algorithm, the computational complexity of the first line in (17) of the revised manuscript is $\mathbf O \big( 2(L+1) \prod^{3}_{k=1} I_k\big)$ and the computational complexity of SVD in the second line is $\mathbf O \big( I^2_kL + L^2 I_k\big)$. Furthermore, the third line consists of a few simple scalar computations, while the fourth line is of complexity order $\mathbf O(L I_k)$. As $L$ is fixed, the computational complexity of (17) at each $t$ is dominated by that of the first step $\mathbf O \big( 2(L+1) \prod^{3}_{k=1} I_k\big)$. Therefore, the computational complexity of the whole algorithm is $\mathbf O \big( 6N_1N_2(L+1) \prod^{3}_{k=1} I_k\big)$, where $N_1$ and $N_2$ are the number of inner and outer iterations, respectively.
 
\section{}
 
 \label{}

  According to (15), when we optimize with respect to
${{{\bm B}^{\left( k \right)}}}$ 
with other variables fixed, 
this subproblem is comprised of a trace norm ${{{\left\| {{{\bm B}^{\left( k \right)}}} \right\|}_ * }}$ (convex) and a positive definite quadratic term ${\frac{1}{2}{{\left\| {\left[ {{{\bm Y}_{\left( k \right)}} - {{\bm B}^{\left( k \right)}}{{\left( {\mathop  \odot \limits_{h \ne k} {{\bm B}^{\left( h \right)}}} \right)}^T}} \right] * {{\bm O}_{\left( k \right)}}} \right\|}^2_F}}$ (strongly convex). Therefore, subproblem (15) is a strongly convex problem.

Let ${\rm{{\cal F}}}\left( {{{\bm B}^{\left( 1 \right)}},{{\bm B}^{\left( 2 \right)}},{{\bm B}^{\left( 3 \right)}}} \right) = \alpha \sum\limits_{k = 1}^3 {{\left\| {{{\bm B}^{\left( k \right)}}} \right\|}_ * } + \frac{1}{2}\bigg\| \bigg[ {\rm{{\cal Y}}} - \sum\limits_{l = 1}^L {\left( {\bm B}_{:,l}^{\left( 1 \right)} \circ {\bm B}_{:,l}^{\left( 2 \right)} \circ {\bm B}_{:,l}^{\left( 3 \right)} \right)}  \bigg] * {\rm{{\cal O}}} \bigg\| ^2_F$. Since ${\rm{{\cal F}}}$ is strongly convex for any single block ${{\bm B}^{\left( k \right)}}$ with $k \in \left\{ {1,2,3} \right\}$, we have~\cite{Wang:20}
\begin{align}     \label{ap1}
		&  {\rm{{\cal F}}}\left( {{{\bm B}^{{{\left( k \right)}^t}}}; {{\bm B}^{{{\left( { - k} \right)}^t}}}} \right) \ge {\rm{{\cal F}}}\left( {{{\bm B}^{{{\left( k \right)}^{t + 1}}}};{{\bm B}^{{{\left( { - k} \right)}^t}}}} \right) + \nonumber \\
	&
	 \left\langle {{\nabla _k}{\rm{{\cal F}}}\left( {{{\bm B}^{{{\left( k \right)}^{t + 1}}}};{{\bm B}^{{{\left( { - k} \right)}^t}}}} \right),{{\bm B}^{{{\left( k \right)}^t}}} - {{\bm B}^{{{\left( k \right)}^{t + 1}}}}} \right\rangle    
	+ \frac{P}{2}{\left\| {{{\bm B}^{{{\left( k \right)}^t}}} - {{\bm B}^{{{\left( k \right)}^{t + 1}}}}} \right\|^2_F},
\end{align}
	for some $P>0$, where ${{\bm B}^{{{\left( { - k} \right)}^t}}} = \big( {{\bm B}^{{{\left( 1 \right)}^{t+1}}}},...,{{\bm B}^{{{\left( {k - 1} \right)}^{t+1}}}},$ ${{\bm B}^{{{\left( {k + 1} \right)}^t}}},...,{{\bm B}^{{{\left( K \right)}^t}}} \big)$ 
and $\left( {{{\bm B}^{{{\left( k \right)}^{t+1}}}};{{\bm B}^{{{\left( { - k} \right)}^t}}}} \right) = \big( {{\bm B}^{{{\left( 1 \right)}^{t+1}}}},...,$ ${{\bm B}^{{{\left( {k - 1} \right)}^{t+1}}}},{{\bm B}^{{{\left( k \right)}^{t+1}}}},{{\bm B}^{{{\left( {k + 1} \right)}^t}}},...,{{\bm B}^{{{\left( K \right)}^t}}} \big)$ with $K = 3$.

Since the subproblem (15) is strongly convex, and the iteration in (17) is monotonic with sufficient decrease property guaranteed after each iteration, the global optimal solution of (15) could be achieved upon convergence of (17) \cite{ABeck}.		
Hence, the first order optimality of (15) holds~\cite{Pavel:21}
	\begin{equation}     \label{ap2}
		\left\langle {{\nabla _k}{\rm{{\cal F}}}\left( {{{\bm B}^{{{\left( k \right)}^{t + 1}}}};{{\bm B}^{{{\left( { - k} \right)}^t}}}} \right),{{\bm B}^{{{\left( k \right)}^t}}} - {{\bm B}^{{{\left( k \right)}^{t + 1}}}}} \right\rangle  \ge 0.
	\end{equation}
Putting (\ref{ap2}) into (\ref{ap1}), we obtain
	\begin{align}    \label{ap3}
		 {\rm{{\cal F}}}\left( {{{\bm B}^{{{\left( k \right)}^t}}};{{\bm B}^{{{\left( { - k} \right)}^t}}}} \right) - & {\rm{{\cal F}}}\left( {{{\bm B}^{{{\left( k \right)}^{t + 1}}}};{{\bm B}^{{{\left( { - k} \right)}^t}}}} \right)    
		 \ge \frac{P}{2}{\left\| {{{\bm B}^{{{\left( k \right)}^t}}} - {{\bm B}^{{{\left( k \right)}^{t + 1}}}}} \right\|^2_F},
	\end{align}
	which is the sufficient decrease condition. Since ${\rm{{\cal F}}}$ is bounded below, taking $t \to \infty $, the left hand side of (\ref{ap3}) must go to zero and we have
	\begin{equation}     \label{ap4}
		\mathop {\lim }\limits_{t \to \infty } \left\| {{{\bm B}^{{{\left( k \right)}^t}}} - {{\bm B}^{{{\left( k \right)}^{t + 1}}}}} \right\|_F = 0,
	\end{equation}
	for any $k \in \left\{ {1,2,3} \right\}$. Now we start to prove the existence of the limit point.

According to \textbf{LEMMA 1}, we know the proposed BCD algorithm is a monotonic algorithm. Start from ${\left\{ {{{\bm B}^{{{\left( k \right)}^0}}}} \right\}_{k = 1,2,3}}$, we have ${\rm{{\cal F}}}\left( {{{\bm B}^{{{\left( 1 \right)}^t}}},{{\bm B}^{{{\left( 2 \right)}^t}}},{{\bm B}^{{{\left( 3 \right)}^t}}}} \right) \le {\rm{{\cal F}}}\left( {{{\bm B}^{{{\left( 1 \right)}^0}}},{{\bm B}^{{{\left( 2 \right)}^0}}},{{\bm B}^{{{\left( 3 \right)}^0}}}} \right)$. Putting back the definition of ${\rm{{\cal F}}}\left( {{{\bm B}^{{{\left( 1 \right)}^t}}},{{\bm B}^{{{\left( 2 \right)}^t}}},{{\bm B}^{{{\left( 3 \right)}^t}}}} \right)$ and eliminate all positive terms except ${\left\| {{{\bm B}^{{{\left( k \right)}^t}}}} \right\|_ * }$, we obtain
	\begin{equation}        \label{ap5}
		{\left\| {{{\bm B}^{{{\left( k \right)}^t}}}} \right\|_ * } \le \frac{1}{\alpha }{\rm{{\cal F}}}\left( {{{\bm B}^{{{\left( 1 \right)}^0}}},{{\bm B}^{{{\left( 2 \right)}^0}}},{{\bm B}^{{{\left( 3 \right)}^0}}}} \right),
	\end{equation}
for any $k \in \left\{ {1,2,3} \right\}$ and $t \ge 0$. Equation (\ref{ap5}) indicates that the sequence $\left\{ {{{\bm B}^{{{\left( k \right)}^{{t}}}}}} \right\}_{k = 1,2,3}^{t > 0}$ generated by the BCD algorthm is restricted in a bounded area. Therefore, Algorithm 1 would have at least one limit point~\cite{Bartle:00}.

If we denote the limit point as ${\left\{ {{{\bm B}^{{{\left( k \right)}^*}}}} \right\}_{k = 1,2,3}}$, there must exists a subsequence $\left\{ {{{\bm B}^{{{\left( k \right)}^{{t_j}}}}}} \right\}_{k = 1,2,3}^{j > 0}$ converges to this limit point~\cite{Bartle:00}. From (\ref{ap4}), we know the subsequence $\left\{ {{{\bm B}^{{{\left( k \right)}^{{t_j + 1}}}}}} \right\}_{k = 1,2,3}^{j > 0}$ also converges to this limit point. Besides,
	since ${{\bm B}^{{{\left( k \right)}^{{t_j} + 1}}}} = \mathop {\arg \min }\limits_{{{\bm B}^{\left( k \right)}}} \left\{ {{\rm{{\cal F}}}\left( {{{\bm B}^{\left( k \right)}};{{\bm B}^{{{\left( { - k} \right)}^{{t_j}}}}}} \right)} \right\}$, we have 
	\begin{equation}     \label{ap6}
		{\rm{{\cal F}}}\left( {{{\bm B}^{{{\left( k \right)}^{{t_j} + 1}}}};{{\bm B}^{{{\left( { - k} \right)}^{{t_j}}}}}} \right) \le {\rm{{\cal F}}}\left( {{{\bm B}^{\left( k \right)}};{{\bm B}^{{{\left( { - k} \right)}^{{t_j}}}}}} \right),
	\end{equation}
	for any ${{{\bm B}^{\left( k \right)}}}$. Then taking $j \to \infty $, since ${\rm{{\cal F}}}$ is a continuous function w.r.t. ${\left\{ {{{\bm B}^{\left( k \right)}}} \right\}_{k = 1,2,3}}$, we obtain
	\begin{equation}   \label{ap7}
		{\rm{{\cal F}}}\left( {{{\bm B}^{{{\left( k \right)}^ * }}};{{\bm B}^{{{\left( { - k} \right)}^ * }}}} \right) \le {\rm{{\cal F}}}\left( {{{\bm B}^{\left( k \right)}};{{\bm B}^{{{\left( { - k} \right)}^ * }}}} \right),
	\end{equation}
for any ${k \in \left\{ {1,2,3} \right\}}$ and ${{{\bm B}^{\left( k \right)}}}$, and equation (\ref{ap7}) means the limit point ${\left\{ {{{\bm B}^{{{\left( k \right)}^*}}}} \right\}_{k = 1,2,3}}$ is a KKT point. Finally, by Theorem 2.8 of~\cite{Xu:13}, since the proposed BCD algorithm achieves the global optimal update in each subproblem, and there is no constraint in (14), the whole sequence ${\left\{ {{{\bm B}^{{{\left( k \right)}^t}}}} \right\}_{k = 1,2,3}}$ converges to ${\left\{ {{{\bm B}^{{{\left( k \right)}^*}}}} \right\}_{k = 1,2,3}}$ as $t$ goes to infinity. Hence, all different limit points induced by different subsequences in ${\left\{ {{{\bm B}^{{{\left( k \right)}^t}}}} \right\}_{k = 1,2,3}}$ converges to ${\left\{ {{{\bm B}^{{{\left( k \right)}^*}}}} \right\}_{k = 1,2,3}}$, which is a unique limit point.

\section{}

\begin{algorithm}
\caption{HaLRTC (noise versioin)}
\label{table:Coal Meta-Heuristic}
{\textbf{Input:} $\mathcal X$, $\rho$, $\gamma$ and $K$;} \\
{\textbf{Output:} $\mathcal X$}
\begin{enumerate}
\item {\textbf{for} $k=0$ to $K$ \\
\textbf{do}}
\item \ \ \ \ {\textbf{for} $i=1$ to $n$ \textbf{do}}
\item \ \ \ \ \ \ \ \ {$\mathcal M_i = \text{fold}_i \left[ \boldsymbol D_{\frac{\alpha_i}{\rho}} \left(\mathcal X_{(i)}+ \frac{1}{\rho}\mathcal Y_{i(i)}\right) \right] $ }
\item \ \ \ \ {\textbf{end for}}
\item \ \ \ \  {$\mathcal X_{i_1,...,i_n}= 
\begin{cases}
\left[ \frac{1}{n} \sum^{n}_{i=1} \left( \mathcal M_i - \frac{1}{\rho} \mathcal Y_i \right) \right]_{i_1,...,i_n}, & \text{if} \ (i_1,...,i_n) \notin 
\Omega \\
\left[ \frac{1}{\rho n + \gamma} \bigg( \gamma \mathcal T + \sum^n_{i=1} ( \rho \mathcal M_i - \mathcal Y_i ) \bigg) \right]_{i_1,...,i_n}, & \text{if} \ (i_1,...,i_n) \in \Omega
\end{cases}$}
\item \ \ \ \  {$\mathcal Y_i=\mathcal Y_i - \rho(\mathcal M_i - \mathcal X) $}
\item {\textbf{end for}}
\end{enumerate}
\end{algorithm}

In order to enable the HaLRTC to handle noisy situation, we extended the HaLRTC by replacing the projection operation with a least squares criterion at the objective function. Having predefined a $\gamma$, the modified HaLRTC formulation is then denoted as:
{
\begin{align}   \label{noiHa} 
&\underset{\mathcal{X},\mathcal M_1,...\mathcal M_{n}   } {\text{min}} \ 
\frac{\gamma}{2} \Vert \mathcal X_{\Omega} - \mathcal T_{\Omega}  \Vert^2_{F}  + 
    \sum^n_{i=1} \alpha_i \Vert  \mathcal M_{i(i)}
                        \Vert_*,  \nonumber \\
& \text{s.t.} \ \mathcal{X}=\mathcal M_i,i=1,...,n   
\end{align}
}\ignorespaces
\noindent
Optimization problem in (\ref{noiHa}) can be seen as a generalized version of noiseless HaLRTC, with a large $\gamma$ represents small noise case and a small $\gamma$ denotes large noise case (when $\gamma$ goes to  infinity, it reduces to the noiseless HaLRTC). Problem (\ref{noiHa}) can still be solved with ADMM as shown in Algorithm E1.

\section{}

\begin{figure*}[t]
  \centering
  \begin{minipage}[b]{0.45\textwidth}
    \includegraphics[width=7cm]{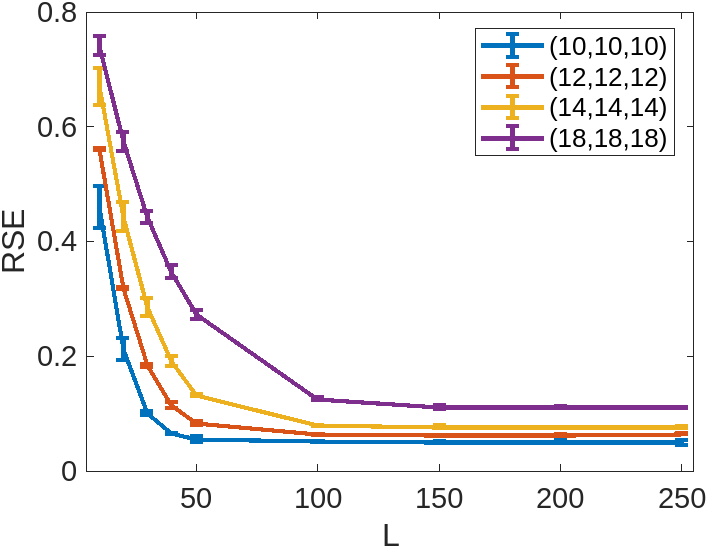}
    \caption{{RSE comparison on synthetic data with SR=20\% and SNR=20dB}}
    \label{fig:figure1}
  \end{minipage}
  \hfill
  \begin{minipage}[b]{0.45\textwidth}
    \includegraphics[width=7cm]{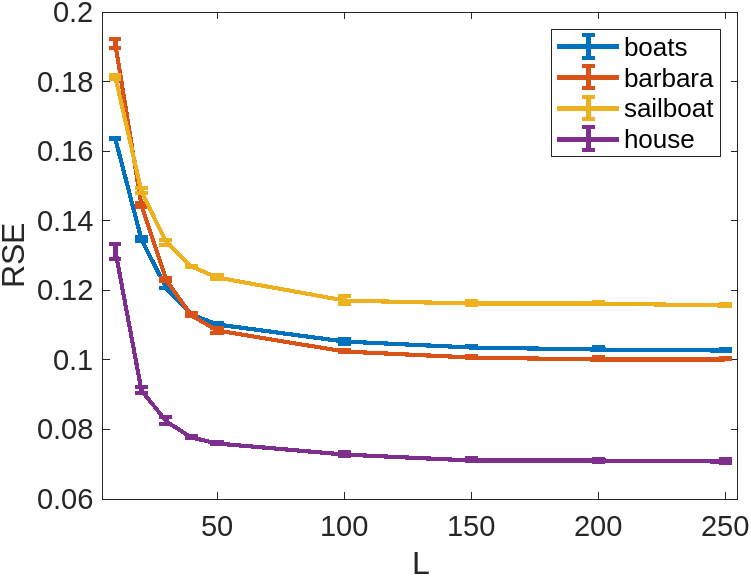}
    \caption{{RSE comparison on color imate with SR=30\% and SNR=20dB}}
    \label{fig:figure2}
  \end{minipage}
\end{figure*}

\begin{figure*}[h]
  \centering
  \begin{minipage}[b]{0.45\textwidth}
    \includegraphics[width=7cm]{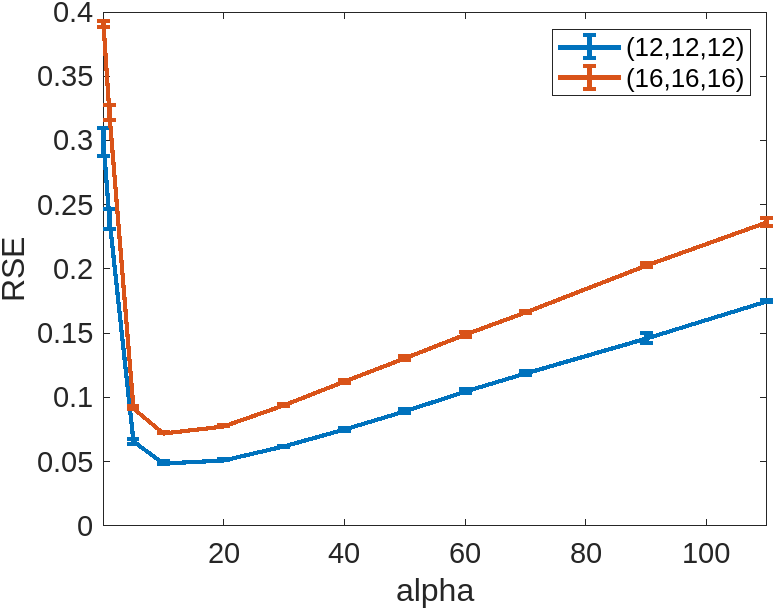}
    \caption{{RSE of synthetic data in SR=20\% and SNR=20dB}}
    \label{fig:figure3}
  \end{minipage}
  \hfill
  \begin{minipage}[b]{0.45\textwidth}
    \includegraphics[width=7cm]{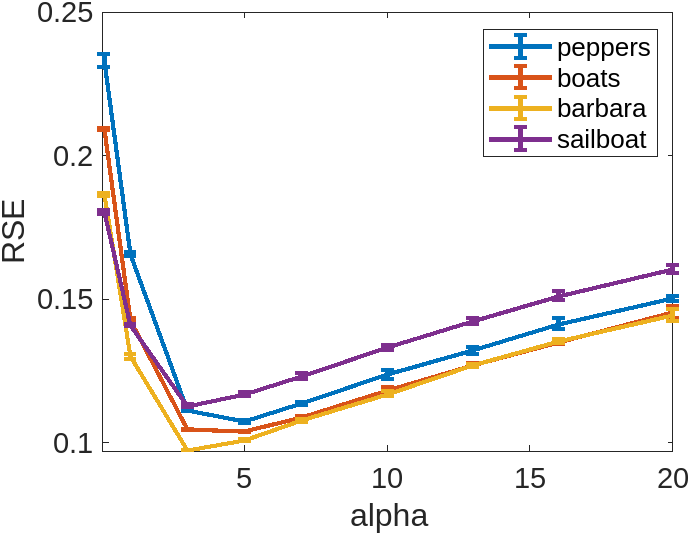}
    \caption{{RSE of 4 color images in SR=30\% and SNR=20dB}}
    \label{fig:figure4}
  \end{minipage}
\end{figure*}

In this Appendix, we assess how the parameters $L$ and $\alpha$ affect the completion performance (with other settings detailed in the manuscript). In the manuscript, we argue from a theoretical 
point-of-view that to accurately represent the Tucker core, the number of columns of $\boldsymbol B^{(k)}$ should be large. To show that this is the case in simulation, we present Figure \ref{fig:figure1} and Figure \ref{fig:figure2} to show the tensor completion performance under different $L$. It can be seen that, in general, the larger the $L$ the better the performance.
Furthermore, when $L$ is greater than 150, the performance is stable and does not have significant improvement.

On the other hand, for the parameter $\alpha$, it represents the trade-off between the fidelity term and the nuclear norm regularization. In general, this parameter needs to be tuned, which is a common practice for optimization-based algorithms. Figure \ref{fig:figure3} and Figure \ref{fig:figure4} show the performance of the proposed algorithm with different $\alpha$. 
It can be seen that performance highly depends on the chosen $\alpha$. Only when $\alpha$ between $\left[5,30\right]$ in synthetic data and $\left[3,5\right]$ in image data, the proposed method delivers good performance.

\begin{figure*}[t]
\renewcommand\arraystretch{0.1}
\centering
\scalebox{0.73}{
\begin{tabular}{@{} m{0.3 em}  m{3.2em} m{3.2em} m{3.2em} m{3.2em} m{3.2em} m{3.2em} m{3.2em} m{3.2em} m{3.2em} m{3.2em} m{3.2em} m{3.2em} @{}}
\scriptsize { } & \scriptsize Original Image & \scriptsize Incomplete Image  & \scriptsize LRFMTC (proposed) & \scriptsize CTNM  & \scriptsize PLM & \scriptsize NNCP & \scriptsize SiLRTC & \scriptsize HaLRTD    & \scriptsize TREL1+W-Tucker  & \scriptsize  LRTV   & \scriptsize McAlm & \scriptsize {KBR} \\
\begin{tabular}{@{}l@{}} \rotatebox{70}{\scriptsize Barbara}  
\end{tabular} &
  \includegraphics[width=.105\textwidth,keepaspectratio]{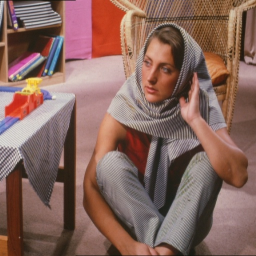}&  
  \includegraphics[width=.105\textwidth]{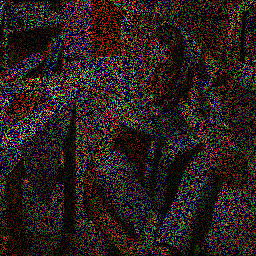}&
  \includegraphics[width=.105\textwidth]{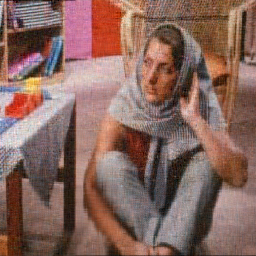}&
  \includegraphics[width=.105\textwidth]{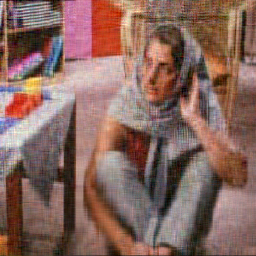}& 
  \includegraphics[width=.105\textwidth]{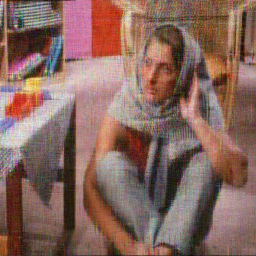}&
     \includegraphics[width=.105\textwidth]{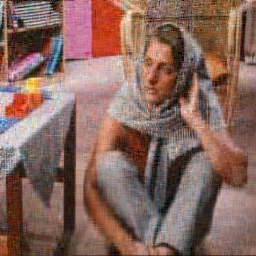}&
  \includegraphics[width=.105\textwidth]{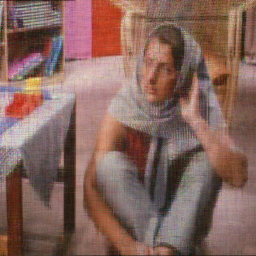}&
  \includegraphics[width=.105\textwidth]{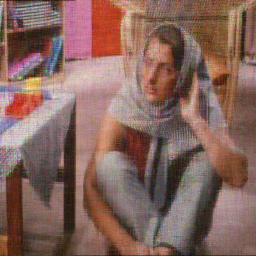}&
  \includegraphics[width=.105\textwidth]{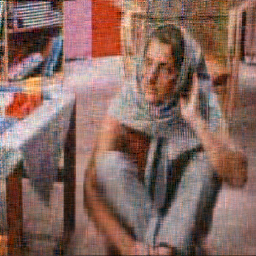}&
  \includegraphics[width=.105\textwidth]{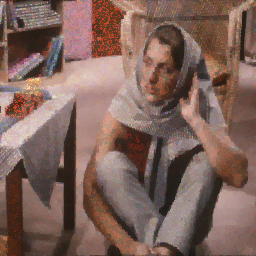}&
  \includegraphics[width=.105\textwidth]{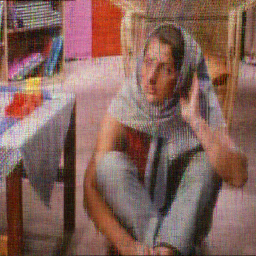}&
  \includegraphics[width=.105\textwidth]{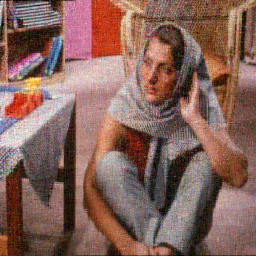}
  \\
\begin{tabular}{@{}l@{}} \rotatebox{70}{\scriptsize Details}  
\end{tabular} &
  \includegraphics[width=.105\textwidth,keepaspectratio]{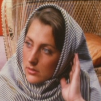}&  
  \includegraphics[width=.105\textwidth]{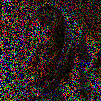}&
  \includegraphics[width=.105\textwidth]{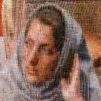}&
  \includegraphics[width=.105\textwidth]{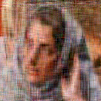}& 
  \includegraphics[width=.105\textwidth]{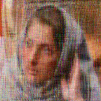}&
   \includegraphics[width=.105\textwidth]{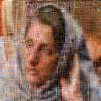}&
  \includegraphics[width=.105\textwidth]{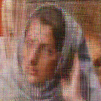}&
  \includegraphics[width=.105\textwidth]{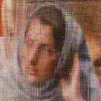}&  
  \includegraphics[width=.105\textwidth]{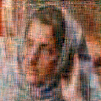}&
  \includegraphics[width=.105\textwidth]{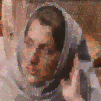}&
  \includegraphics[width=.105\textwidth]{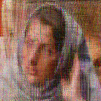}&
  \includegraphics[width=.105\textwidth]{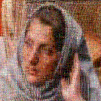}
  \\ 
 \begin{tabular}{@{}l@{}} \rotatebox{70}{\scriptsize Sailboat} 
\end{tabular}  &
  \includegraphics[width=.105\textwidth,keepaspectratio]{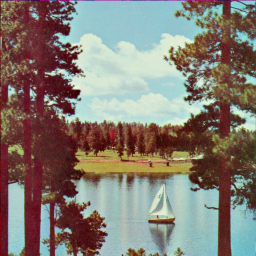}&  
  \includegraphics[width=.105\textwidth]{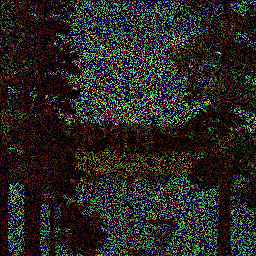}&
  \includegraphics[width=.105\textwidth]{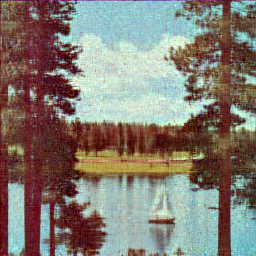}&
  \includegraphics[width=.105\textwidth]{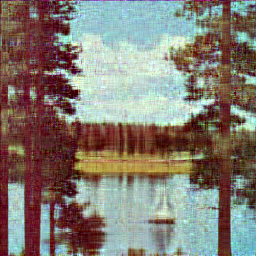}& 
  \includegraphics[width=.105\textwidth]{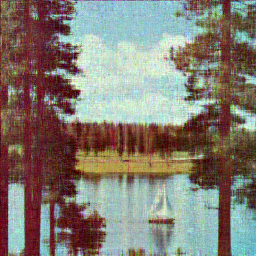}&
    \includegraphics[width=.105\textwidth]{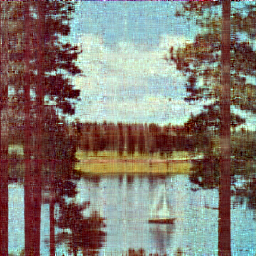}&
  \includegraphics[width=.105\textwidth]{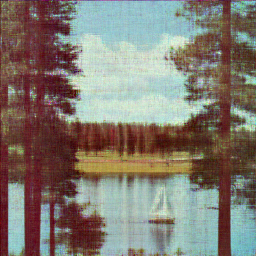}&
  \includegraphics[width=.105\textwidth]{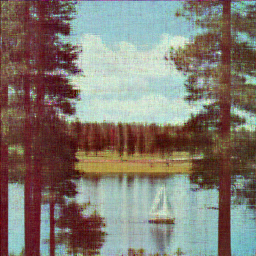}&
  \includegraphics[width=.105\textwidth]{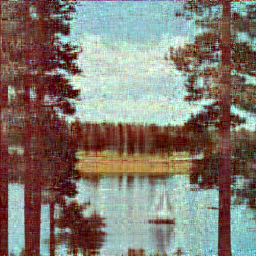}&
  \includegraphics[width=.105\textwidth]{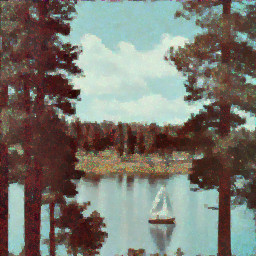}&
  \includegraphics[width=.105\textwidth]{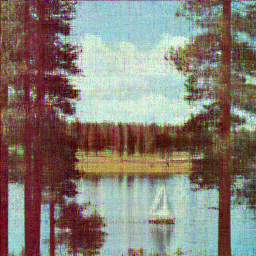}&
  \includegraphics[width=.105\textwidth]{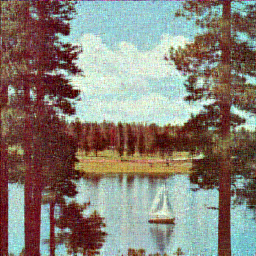}
  \\
\begin{tabular}{@{}l@{}} \rotatebox{70}{\scriptsize Details}  
\end{tabular}  &
  \includegraphics[width=.105\textwidth,keepaspectratio]{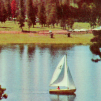}&  
  \includegraphics[width=.105\textwidth]{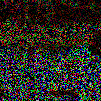}&
  \includegraphics[width=.105\textwidth]{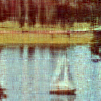}&
  \includegraphics[width=.105\textwidth]{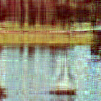}& 
  \includegraphics[width=.105\textwidth]{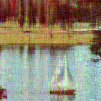}&
    \includegraphics[width=.105\textwidth]{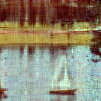}&
  \includegraphics[width=.105\textwidth]{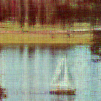}&
  \includegraphics[width=.105\textwidth]{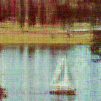}&  
  \includegraphics[width=.105\textwidth]{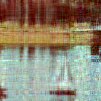}&
  \includegraphics[width=.105\textwidth]{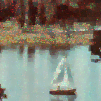}&
  \includegraphics[width=.105\textwidth]{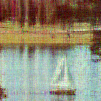}&
  \includegraphics[width=.105\textwidth]{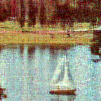}\\
  
   \begin{tabular}{@{}l@{}} \rotatebox{70}{\scriptsize Peppers} 
\end{tabular}  &
  \includegraphics[width=.105\textwidth,keepaspectratio]{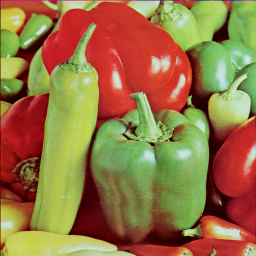}&  
  \includegraphics[width=.105\textwidth]{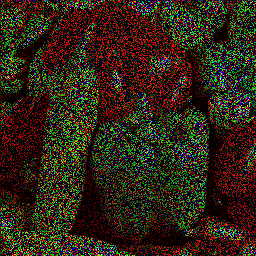}&
  \includegraphics[width=.105\textwidth]{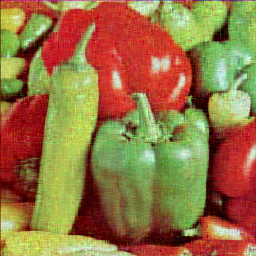}&
  \includegraphics[width=.105\textwidth]{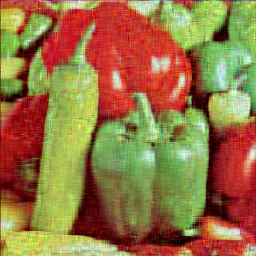}& 
  \includegraphics[width=.105\textwidth]{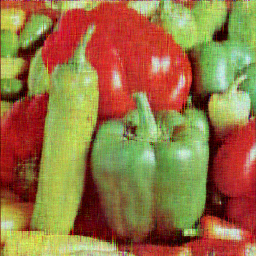}&
    \includegraphics[width=.105\textwidth]{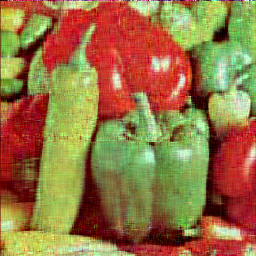}&
  \includegraphics[width=.105\textwidth]{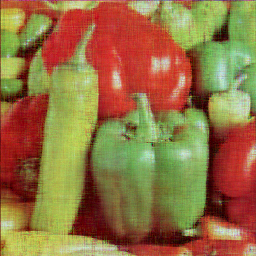}&
  \includegraphics[width=.105\textwidth]{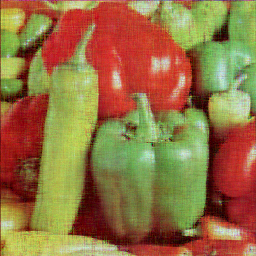}&
  \includegraphics[width=.105\textwidth]{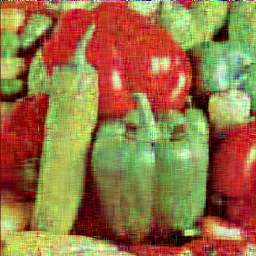}&
  \includegraphics[width=.105\textwidth]{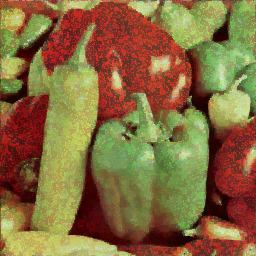}&
  \includegraphics[width=.105\textwidth]{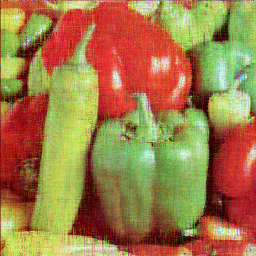}&
  \includegraphics[width=.105\textwidth]{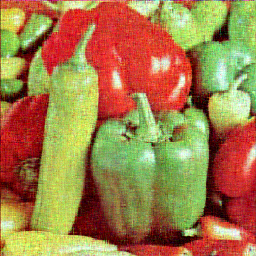}\\  
\begin{tabular}{@{}l@{}}   \rotatebox{70}{\scriptsize Details}  
\end{tabular}  &
  \includegraphics[width=.105\textwidth,keepaspectratio]{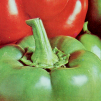}&  
  \includegraphics[width=.105\textwidth]{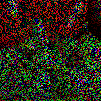}&
  \includegraphics[width=.105\textwidth]{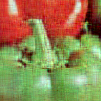}&
  \includegraphics[width=.105\textwidth]{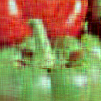}& 
  \includegraphics[width=.105\textwidth]{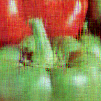}&
    \includegraphics[width=.105\textwidth]{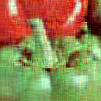}&
  \includegraphics[width=.105\textwidth]{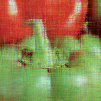}&
  \includegraphics[width=.105\textwidth]{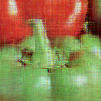}&  
  \includegraphics[width=.105\textwidth]{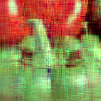}&
  \includegraphics[width=.105\textwidth]{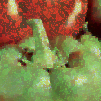}&
  \includegraphics[width=.105\textwidth]{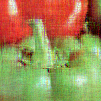}&
  \includegraphics[width=.105\textwidth]{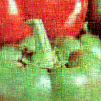}\\
  
  \begin{tabular}{@{}l@{}}  \rotatebox{70}{\scriptsize Boats} 
\end{tabular}  &
  \includegraphics[width=.105\textwidth,keepaspectratio]{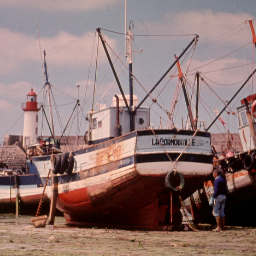}&  
  \includegraphics[width=.105\textwidth]{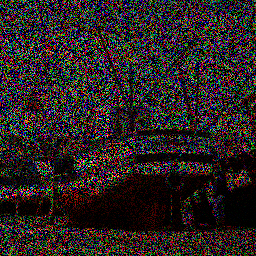}&
  \includegraphics[width=.105\textwidth]{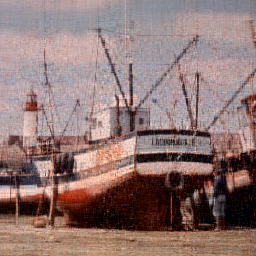}&
  \includegraphics[width=.105\textwidth]{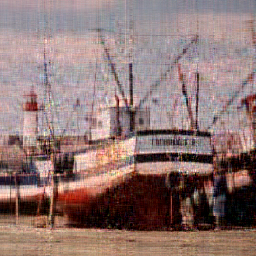}& 
  \includegraphics[width=.105\textwidth]{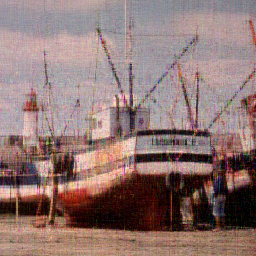}&
    \includegraphics[width=.105\textwidth]{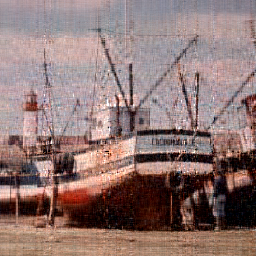}&
  \includegraphics[width=.105\textwidth]{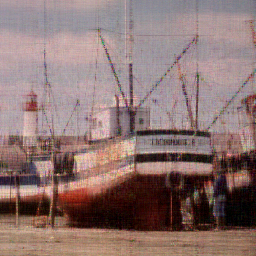}&
  \includegraphics[width=.105\textwidth]{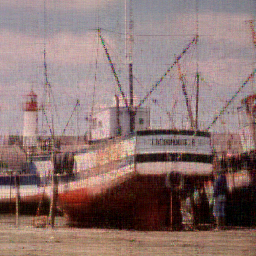}&
  \includegraphics[width=.105\textwidth]{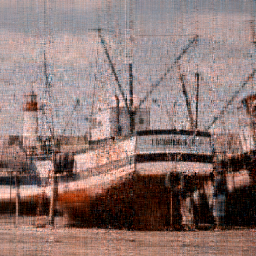}&
  \includegraphics[width=.105\textwidth]{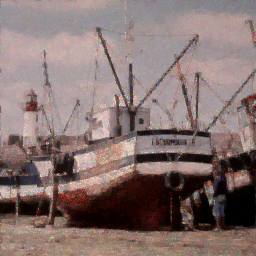}&
  \includegraphics[width=.105\textwidth]{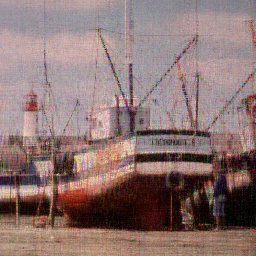}&
  \includegraphics[width=.105\textwidth]{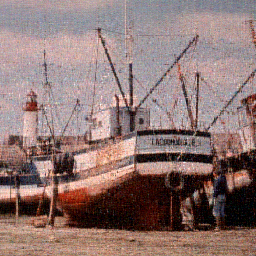}\\
  

\begin{tabular}{@{}l@{}}   \rotatebox{70}{\scriptsize Details}  
\end{tabular}  &
  \includegraphics[width=.105\textwidth,keepaspectratio]{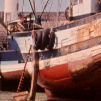}&  
  \includegraphics[width=.105\textwidth]{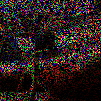}&
  \includegraphics[width=.105\textwidth]{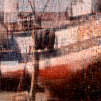}&
  \includegraphics[width=.105\textwidth]{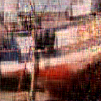}& 
  \includegraphics[width=.105\textwidth]{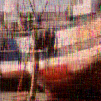}&
    \includegraphics[width=.105\textwidth]{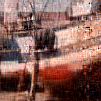}&
  \includegraphics[width=.105\textwidth]{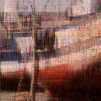}&
  \includegraphics[width=.105\textwidth]{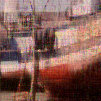}&  
  \includegraphics[width=.105\textwidth]{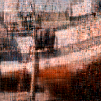}&
  \includegraphics[width=.105\textwidth]{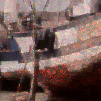}&
  \includegraphics[width=.105\textwidth]{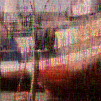}&
  \includegraphics[width=.105\textwidth]{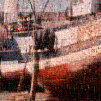}\\
    \end{tabular}
    }
\captionsetup{font={footnotesize,stretch=1}}
\caption{Examples of the recovered images at SR=30\% and SNR=20dB.}
\label{SampleImages}
\end{figure*}

\section{}

To examine the visual differences, Figure \ref{SampleImages} shows several examples (Barbara, Sailboat, Peppers and Boats) of reconstructed images from all the compared algorithms (locally enlarged details are shown in the second, forth, sixth and eighth rows). It can be seen that the proposed LRFMTC recovers the best images, achieving a better balance between retaining image details and noise removal. Compared to the proposed LRFMTC, CTNM loses more image details, PLM gains more noise in vertical direction, SiLRTC and HaLRTC have less expressive ability for both the color saturation and image details, NNCP is relatively weak in noise removal and TREL1+W-Tucker has weaker power for not only noise removal, but also details recovery.

\begin{figure*}[t]
\centering
\subfigure[]{
\begin{minipage}[t]{0.307\textwidth}
\centering
\includegraphics[width=5.0cm]{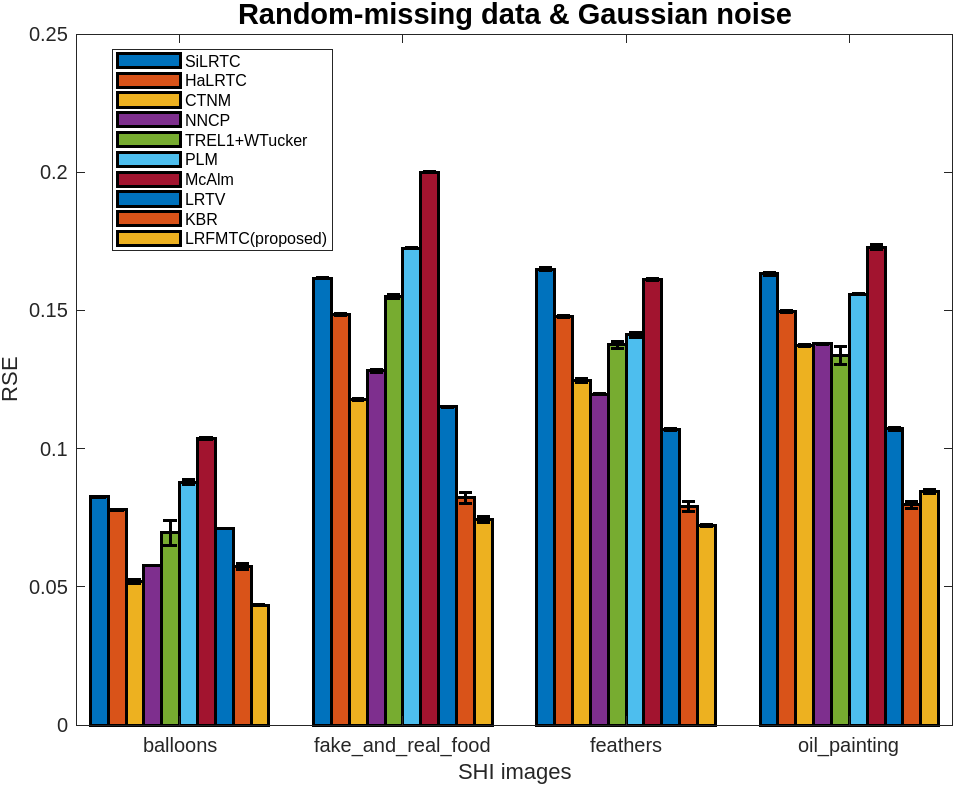}
\end{minipage}
}
\subfigure[]{
\begin{minipage}[t]{0.307\textwidth}
\centering
\includegraphics[width=5.0cm]{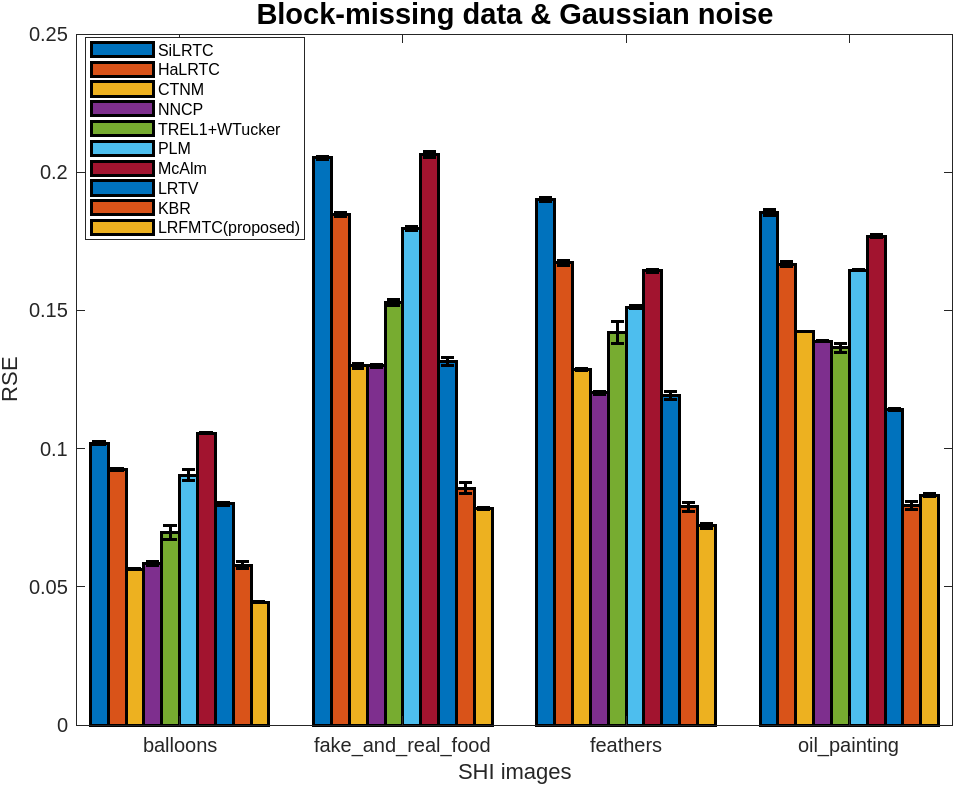}
\end{minipage}
}
\subfigure[]{
\begin{minipage}[t]{0.307\textwidth}
\centering
\includegraphics[width=5.0cm]{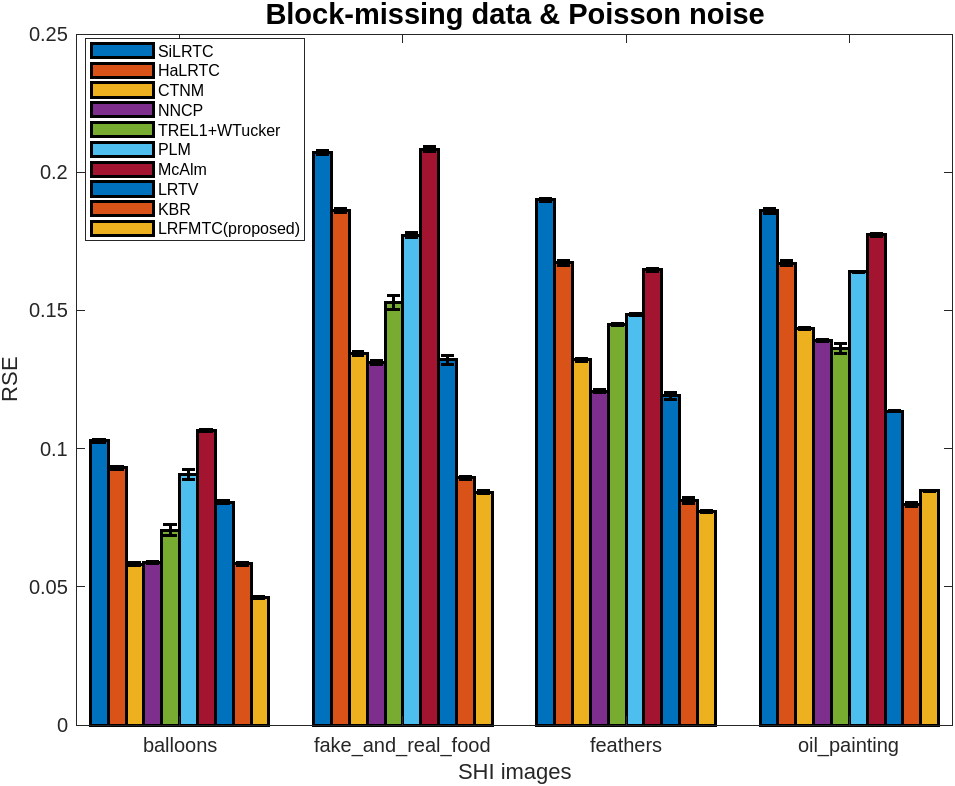}
\end{minipage}
}
\centering
\captionsetup{font={footnotesize,stretch=1}}
\caption{Averaged RSE comparison under a) Random-missing data \& Gaussian noise b) Block-missing data \& Gaussian noise and  c) Block-missing data \& Poisson noise on 4 HSI images (SR=0.3,SNR=20dB). The vertical error bars show one standard deviation.}
\label{HSIRSE}
\end{figure*}

\begin{table*}[t]
\caption{PSNR and SSIM comparison in HyperSpectral Image (HSI) completion (SR=0.3) with 20dB signal-to-noise ratio}
\footnotesize
\centering
\scalebox{0.7}{
\begin{tabular}{ l |m{2em} m{2em} |m{2em} m{2em} |m{2em} m{2em} |m{2em} m{2em}|m{2em} m{2em} |m{2em} m{2em} |m{2em} m{2em} |m{2em} m{2em} |m{2em} m{2em}|m{2em} m{2em} }
 \hline
 \quad 
 & \multicolumn{2}{c|}{SiLRTC} & \multicolumn{2}{c|}{{HaLRTC}} & \multicolumn{2}{c|}{CTNM} & \multicolumn{2}{c|}{NNCP} & \multicolumn{2}{c|}{\begin{tabular}{@{}c@{}} TREL1+ \\ WTucker\end{tabular}} & \multicolumn{2}{c|}{PLM} & \multicolumn{2}{c|}{McAlm}    & \multicolumn{2}{c|}{LRTV} & 
 \multicolumn{2}{c|}{{KBR}} & \multicolumn{2}{c}{\begin{tabular}{@{}c@{}} (Proposed) \\ LRFMTC\end{tabular}}\\
\textbf{} & PSNR & SSIM  &  PSNR &  SSIM &  PSNR &  SSIM &  PSNR &  SSIM & PSNR & SSIM &  PSNR &  SSIM &  PSNR &  SSIM &  PSNR &  SSIM & PSNR & SSIM  & PSNR & SSIM \\ \hline    
\bf{Part I}: &  \multicolumn{16}{c}{\bf{Random-missing data \& Gaussian noise} }      \\   \hline
balloons   & 35.15   & 0.952  & {35.72}  &  {0.958}  & 39.28  &  0.974  & 38.37  &  0.969  & 36.72  &  0.953  & 34.69  &  0.930  & 33.19  &  0.920 &36.49    &0.952 &{38.28}&{0.943} & \bf{40.85}  &  \bf{0.979 }\\
fake\&real\_food &  33.13  &  0.931 & {34.25}  &  {0.940} &  36.09   & 0.950  & 35.35  &  0.939  & 33.69  &  0.913   &32.71  &  0.907 &  31.32  &  0.896 &  36.17  &  0.956 &{39.03}&{0.949} & \bf{40.07}  &  \bf{0.973}   \\
feathers        &     28.88  &  0.900  & {29.97}   & {0.911} &  31.54   & 0.912  & 31.86   & 0.906 &  30.67  &  0.874  & 30.42  &  0.869 &  29.13 &   0.874 &  32.74  &  0.922 &{35.32}&{0.908} & \bf{36.21}&   \bf{0.952} \\ 
oil\_painting     &    29.53   & 0.882  & {30.39}  &  {0.901}  & 31.08  &  0.901  & 31.02  &  0.901  & 31.29  &  0.901 &  29.95 &   0.874  & 29.05 &   0.864   &33.22 &   0.935  &{\bf 35.78} & {0.946}& {35.28}  &  \bf{0.954}
\\ \hline
\bf{Part II}: &  \multicolumn{16}{c}{\bf{Block-missing data \& Gaussian noise} }      \\   \hline
balloons        &     33.30 &   0.939   & {34.26}   &  {0.949}   & 38.61   &  0.969  &  38.27   &  0.968   & 36.73   &  0.953   & 34.45   &  0.928   & 33.03    & 0.917   & 35.46   &  0.947  &{38.23}&{0.944} & \bf{40.61 }   & \bf{0.978}
      \\
fake\&real\_food &   31.04  &  0.906  & {32.55} &   {0.923} & 35.31   & 0.944 &  35.24 &   0.938   &33.82  &  0.915  & 32.37  &  0.900  & 31.05  &  0.890  & 34.99 &   0.949  &{38.70}&{0.948}& \bf{39.62} &   \bf{0.970}
\\
feathers        &      27.64   & 0.882  & {28.98}  &  {0.897}  & 31.30  &  0.907  & 31.81   & 0.905 &  30.34 &   0.877  & 29.81   & 0.863  & 28.95  &  0.872 &  31.76  &  0.910   &{35.32}&{0.908}&\bf{36.23}   & \bf{0.951}
      \\
oil\_painting         &   28.42  &  0.858 &  {29.46}  &  {0.885}  & 30.79  &  0.895 &  30.98 &   0.899 &  31.12  &  0.899 &  29.49 &   0.862  & 28.85  &  0.859 &  32.67 &   0.928    &{\bf35.81}&{0.946} &{35.42}  &  \bf{0.955}
\\ \hline
\bf{Part III}: &  \multicolumn{16}{c}{\bf{Block-missing data \& Poisson noise} }      \\   \hline
balloons        &      33.21 &  0.946 &  {34.17}  &  {0.956}  & 38.32  &  0.969 &  38.20  &  0.968 &  36.60  &  0.952 &  34.36 &   0.940  & 32.92 &   0.931  & 35.34 &   0.958 &{38.17}&{0.957} &\bf{40.30}  &  \bf{0.978}
     \\
fake\&real\_food &    30.95  &  0.914  & {32.53}  &  {0.931}  & 35.01  &  0.942  & 35.16 &   0.937  & 33.82  &  0.916  & 32.30  &  0.919 &  30.92   & 0.906  & 34.85 &  0.964 &{38.53}&{0.969} & \bf{38.95}   & \bf{0.971}
\\
feathers       &     27.63   &  0.900   & {28.95}  &   {0.913}   & 31.06   &  0.906  &  31.77    & 0.906   & 30.17   &  0.876    &29.78   &  0.904   & 28.88   &  0.902  &  31.69   &  0.935 &{35.10}&{0.941}  & \bf{35.60}  &   \bf{0.952}
      \\
oil\_painting         &      28.39    &  0.863   &  {29.44}   &  {0.890}    & 30.72    &  0.895    & 30.95    &  0.899     &31.13    &  0.900   &  29.47    &  0.873    & 28.81    &  0.868   &  32.67     & 0.936     &{\bf35.74}&{0.951} &{35.25  }    &\bf{0.955}
     \\
\hline
\end{tabular}}
\label{HSIPSNRSSIM}
\end{table*}

\section{}

We used 4 hyperspectral images (HSIs) chosen from the CAVE dataset\footnote{{http://www1.cs.columbia.edu/CAVE/databases/multispectral/.}} as test data and evaluated the proposed method and all comparison methods at SR=30\% and SNR=20dB. The results are shown in Figure \ref{HSIRSE} and Table \ref{HSIPSNRSSIM}. In particular, for random missing data patterns and Gaussian noise, the results are presented in Figure \ref{HSIRSE}(a) and the first part of Table \ref{HSIPSNRSSIM}.  To assess the performance beyond random missing data, we also simulated  
the block-wise missing pattern according to the $l$-tuple missingness \cite{l_tuple} with $l=4$ and the results are shown in Figure \ref{HSIRSE} (b) and Part II of Table \ref{HSIPSNRSSIM}.
Furthermore, to evaluate the performance beyond Gaussian noise, we have added the Poisson noise (the lighter parts of the images are noisier than the darker parts) under block-wise missing pattern in the experiments and the results are shown in Figure \ref{HSIRSE} (c) and Part III of Table \ref{HSIPSNRSSIM}.
From Figure \ref{HSIRSE} and Table \ref{HSIPSNRSSIM}, it can be seen that the proposed method performs the best in various criteria (RSE, PSNR, and SSIM).

 \end{appendices}

\end{document}